\documentclass[elsarticle.cls,superscriptaddress, notitlepage]{revtex4-1}
\usepackage{amsmath,amssymb,amsthm}
\usepackage{braket}
\usepackage{times,txfonts}
\usepackage{braket}
\usepackage{color}
\usepackage{mathtools}
\usepackage{natbib}
\usepackage{latexsym}
\usepackage{tabularx, booktabs}
\usepackage{amssymb}
\usepackage{graphics,epstopdf}
\usepackage{graphicx}
\usepackage[colorlinks=true, citecolor=blue, urlcolor=blue ]{hyperref}
\usepackage{float}
\usepackage{graphicx}
\usepackage{amsfonts}
\usepackage{color}
\usepackage{caption}
\usepackage{subcaption}
\usepackage{wrapfig, framed, caption}

\begin{document}
	
\title{Characterization of the quantumness of unsteerable tripartite correlations}

\author{Debarshi Das}
\email{debarshidas@jcbose.ac.in}
\affiliation{Centre for Astroparticle Physics and Space Science (CAPSS), Bose Institute, Block EN, Sector V, Salt Lake, Kolkata 700 091, India}

\author{C. Jebaratnam}
\email{jebarathinam@bose.res.in}
\affiliation{S. N. Bose National Centre for Basic Sciences, Block JD, Sector III, Salt Lake, Kolkata 700 098, India}

\author{Bihalan Bhattacharya}
\email{bihalan@gmail.com}
\affiliation{S. N. Bose National Centre for Basic Sciences, Block JD, Sector III, Salt Lake, Kolkata 700 098, India}

\author{Amit Mukherjee}
\email{amitisiphys@gmail.com}
\affiliation{Optics and Quantum Information Group, 
The Institute of Mathematical Sciences, HBNI, C. I. T. Campus, Taramani, Chennai 600 113, 
India}

\author{Some Sankar Bhattacharya}
\email{somesankar@gmail.com}
\affiliation{Physics and Applied Mathematics Unit, Indian Statistical Institute, 203 B. T. Road, Kolkata 700 108, India}
\affiliation{Department of Computer Science, The University of Hong Kong, Pokfulam Road, Hong 
Kong}

\author{Arup Roy}
\email{arup145.roy@gmail.com}
\affiliation{Physics and Applied Mathematics Unit, Indian Statistical Institute, 203 B. T. Road, Kolkata 700 108, India}

\begin{abstract}
	
Quantumness for a bipartite unsteerable quantum correlation is operationally characterized by the notion of super-unsteerability. Super-unsteerability refers
to the requirement of a larger dimension of the random variable that the steering party has to preshare with the party to be steered in the
classical simulation protocol to generate an unsteerable correlation than  the local Hilbert space dimension of  the quantum  states (reproducing the given unsteerable correlation) at the steering party's side. In the present study, this concept of super-unsteerability is generalized by defining the notion of super-bi-unsteerability for tripartite correlations, which is unsteerable across a bipartite cut. Genuine super-bi-unsteerability is defined as the occurrence of super-bi-unsteerability across all possible bipartite cuts. Specific example of genuine super-bi-unsteerability for tripartite correlations has been presented. This study provides a tool to characterize the genuine quantumness of tripartite quantum correlations which are unsteerable across every bipartite cut.

\end{abstract} 
	
	
	\maketitle
	
\section{ INTRODUCTION} 
Quantum composite systems exhibit several nonclassical features such as  entanglement \cite{ent}, Einstein-Podolsky-Rosen (EPR) steering \cite{epr, steer, steer2} and Bell nonlocality \cite{Bell,chsh, bell2}. In the Bell scenario, local quantum measurements on certain spatially separated system leads to nonlocal correlations which cannot be explained by local hidden variable (LHV) theory \cite{Bell}. However, it is well-known that quantum mechanics (QM) is not maximally nonlocal as there are post-quantum correlations, obeying the no-signalling (NS) principle, which are more nonlocal than QM. Popescu-Rohrlich (PR) box \cite{pr} is one such correlation. Nonlocality in QM is limited by the Tsirelson bound \cite{tsi}.

Motivated by the seminal argument by Einstein, Podolsky and Rosen (EPR) \cite{epr}  demonstrating the incompleteness of QM, Schrodinger introduced the concept of `quantum steering' \cite{scro}. The task of quantum steering \cite{steer, steer2} is to prepare different ensembles at one part of a bipartite system by performing local quantum measurements on another part of the bipartite system in such a way that these ensembles cannot be explained by a local hidden state (LHS) model. In other words, quantum correlations, which are steerable, cannot be reproduced by local hidden variable-local hidden state (LHV-LHS) model. In recent years, studies related to quantum steering have been acquiring considerable interest, as witnessed by a wide range of studies \cite{st8, steer22, steer3, st4, st9, st5, steer24, s24}.
Bell-nonlocal states form a subset of the steerable states which also form a subset of the entangled states \cite{steer, st11}. However, unlike quantum nonlocality and entanglement, the task of quantum steering is inherently asymmetric \cite{st7}. In this case, the outcome statistics of one subsystem (which is being ‘steered’) is due to valid QM measurements on a valid QM state. On the other hand, there is no such constraint for the other subsystem.  Quantum steering has also applications in semi device independent scenario where the party, which is being steered, has trust on his/her quantum device but the other party's device is untrusted. Secure quantum key distribution (QKD) using quantum steering has
been demonstrated \cite{stqkd}, where one party cannot trust his/her
devices.

Recently, it has been demonstrated that certain quantum
information tasks may become advantageous even using separable states if they have quantum discord \cite{disc, disc2, disc3}, which is a generalized measure of quantum correlations. This motivated the study of nonclassicality going beyond nonlocality. Certain separable states which have 
quantumness may improve quantum protocols if the shared randomness between the parties is finite \cite{BP14}. This provides an operational
meaning of the measures of quantumness such as quantum discord.
In the context of classical simulation of local entangled states, 
Bowles et. al. \cite{BHQ+15} have shown that the statistics of all local entangled states can be simulated by using only finite shared randomness and they defined a measure which is the minimal dimension of that shared classical
randomness. On the other hand, all the previous works have used unbounded shared randomness to simulate a given local entangled state. In Ref. \cite{sl1}, the minimal dimension of the shared classical randomness
required to simulate any local correlation in a given Bell scenario have been demonstrated. Motivated by this, an interesting feature of certain local boxes, called superlocality, has been defined as follows:
there exist certain local boxes which can be simulated by quantum systems of local dimension lower than the minimum dimension of the shared classical randomness needed to simulate them. This implies that superlocality refers to the dimensional advantage in simulating certain local boxes by using quantum systems. In particular, it has been shown \cite{sl1, sl2} that entanglement enables superlocality, however, superlocality occurs even for separable states. Recently it has been pointed out \cite{sl5} that superlocality cannot occur for arbitrary separable states, in particular, the
separable states which are a classical-quantum state \cite{bc} or its permutation can never lead to superlocality. One important point to be stressed here is that the bipartite quantum states which are not a classical-quantum state must have quantumness as quantified by quantum discord. Recently, Generalizing the concept of superlocality, the notion of super-correlation \cite{sus} has been defined as follows: the requirement for a larger dimension of the preshared
randomness to simulate the correlations than that of the quantum states that generate them. In particular, the quantumness of certain unsteerable correlations has been pointed out by the notion of super-unsteerability \cite{sus}, the requirement for a larger
dimension of the classical variable that the steering party has to preshare with the party to be steered for simulating the unsteerable correlation than the local Hilbert space dimension of the quantum system (reproducing the given unsteerable correlation) at the steering party's side (i. e., at the untrusted party's side).

The extension of the Bell-type scenario to more than two parties was first presented in the seminal work by  Greenberger, Horne, and Zeilinger \cite{GHZ}. Certain interesting features of nonlocality in tripartite scenario have been established \cite{SI,mermin,PM1,PM2}.  Genuine tripartite quantum discord has been defined to quantify the quantumness shared among all three subsystems of the tripartite quantum state \cite{mp1, mp2, mp3}. Recently, it has been demonstrated that the limited dimensional quantum simulation of certain local  tripartite correlations must require genuine tripartite quantum discord states. To study genuine nonclassicality of these correlations, two quantities called, Svetlichny strength and Mermin strength has been defined in the context of tripartite NS boxes \cite{mp4}.

In case of multipartite systems, earlier studies have established that certain nonlocal measures may indeed be amplified by the addition of system dimensions \cite{mp5, mp6, mp7, mp8}. Multipartite quantum entanglement displays complicated structures, which can be broadly classified according to whether entanglement is shared among all subsystems of a given multipartite system or not. In this context, the notion of genuine multipartite nonlocality has been introduced and Bell-type inequalities have been derived to detect it \cite{SI}.
Genuine multipartite quantum nonlocality can be quantified by classical communication 
models, where the $n$ parties are grouped into $m$ disjoint groups; within each group, the parties can freely communicate with each other, but are not allowed to do the same between distinct groups \cite{mp9}. The minimal amount of communication between these disjoint groups required to reproduce a given nonlocal correlation determines the extent of multipartite quantum nonlocality of that correlation. Recently, the operational characterization of genuine nonclassicality of local multipartite correlations has been presented and the notion of superlocality has been generalised in the context of local multipartite correlations \cite{mp10}.

 The concept of EPR steering  as formalized in the bipartite scenario in Ref. \cite{steer}
has been generalized for multipartite scenarios in Refs. \cite{UFNL,stm2}. Subsequent to these studies, in Refs. \cite{st10, stm4, stm42},
genuine multipartite steering, in which nonlocality in the form steering is necessarily shared among all observers has been proposed.
In Ref. \cite{st10}, genuine multipartite steering was defined from the foundational perspective, i.e., in terms of the nonexistence of the 
hybrid LHS local-nonlocal model for the measurement correlations between the parties. In Ref. \cite{cava}, quantum information theoretic characterization
of genuine multipartite steering was proposed as the detection of genuine multipartite entanglement in the partially device-independent 
scenarios where some of the parties' measuring devices are trusted while the others are untrusted.
Genuine tripartite steering inequalities have also been derived \cite{cava, stm6, sttr1, sttr2, stm7} to detect genuine tripartite entanglement
in the one-sided and two-sided device-independent scenarios.

Against the above backdrop, the motivation of the present study is to generalize the notion of super-unsteerability in the tripartite scenario to analyze the resource
requirement for simulating the tripartite bi-unsteerable correlations (which are unsteerable across some particular bipartition) in the
context of the steering scenario where one of the parties' measurements are untrusted while the other two parties' 
measurements are trusted.  In particular, we show that quantumness is necessary to reproduce certain tripartite bi-unsteerable correlations in the scenario where the dimension of the resource reproducing the correlations is
restricted. We demonstrate that there are certain tripartite bi-unsteerable
correlations whose simulation with LHV-LHS model requires
preshared randomness with dimension exceeding the local Hilbert space dimension of the quantum system (reproducing the given bi-unsteerable correlation) at the untrusted party's side. This is termed as \textit{``super-bi-unsteerability"} across some particular bipartition. Moreover, we define \textit{``genuine super-bi-unsteerability"} as the occurrence of super-bi-unsteerability across all possible bipartitions. It provides a tool to give an operational characterization of the genuine quantumness of certain tripartite correlations which are bi-unsteerable across all possible bipartitions.

The plan of the paper is as follows. In Section II, the basic notions of NS
polytope and the fundamental ideas of quantum steering in bipartite and tripartite scenario has been presented. Our purpose is to decompose the given NS correlation in terms of convex combinations of extremal boxes of NS polytope which leads to a LHV-LHS decomposition of the given correlation.  In Section III, we demonstrate the formal definition of super-bi-unsteerability as well as genuine super-bi-unsteerability, which is followed by Section IV presenting specific examples of genuine super-bi-unsteerability. In Section V,  genuine quantumness of certain bi-unsteerable tripartite correlations captured by genuine super-bi-unsteerability has been discussed. Finally, in the concluding Section VI, we elaborate a bit on the significance of the results obtained.

\section{Framework}

\subsection{No-signalling Polytope}

Let us consider the quantum correlations arising from the following tripartite Bell scenario. Suppose, three spatially separated parties (say, Alice, Bob and Charlie) share a quantum mechanical system $\rho_{ABC} \in \mathcal{B}( \mathcal{H}_A \otimes \mathcal{H}_B \otimes \mathcal{H}_C)$, where $\mathcal{H}_K$ denotes Hilbert space of $k$th party and $\mathcal{B}( \mathcal{H}_A \otimes \mathcal{H}_B \otimes \mathcal{H}_C)$ stands for the set of all bounded linear operators acting on the Hilbert space $\mathcal{H}_A \otimes \mathcal{H}_B \otimes \mathcal{H}_C$.
In this scenario, a correlation between the outcomes is described by the set of conditional
probability distributions $P(abc|A_x B_y C_z)$, where $A_x$, $B_y$, and $C_z$ denote the inputs (measurement choices) 
and $a$, $b$ and $c$ denote the outputs (measurement outcomes) of Alice, Bob and Charlie respectively 
(with $x,y,z,a,b,c\in \{0,1\}$).
Suppose $M^{a}_{A_x}$,  $M^{b}_{B_y}$ and $M^{c}_{C_z}$ denote the measurement operators
of Alice, Bob, and Charlie, respectively.
Then any conditional probability distribution can be expressed in quantum mechanics through the Born's rule as follows:
\begin{equation}
P(abc|A_x B_y C_z)=\mathrm{Tr} \left(\rho_{ABC} M^{a}_{A_x}\otimes M^{b}_{B_y}\otimes M^{c}_{C_z}\right).
\end{equation}
%In particular, we focus on the scenario where the parties share a three-qubit state  
%and perform spin projective measurements corresponding to the operators: $A_x=\hat{a}_x.\vec{\sigma}$, $B_y=\hat{b}_y.\vec{\sigma}$, and $C_z=\hat{c}_z.\vec{\sigma}$. Here $\hat{a}_x$, $\hat{b}_y$, and $\hat{c}_z$ are unit Bloch vectors denoting the measurement directions and $\vec{\sigma}=\{\sigma_1,\sigma_2,\sigma_3\}$, with $\{\sigma_i\}_{i=1,2,3}$ being the Pauli matrices.\\

The set of no-signaling (NS) boxes with two binary inputs and two binary outputs
forms a convex polytope $\mathcal{N}$ in a $26$ dimensional space \cite{ns1}, which includes the set of quantum correlations $Q$ as a proper subset. 
Any box belonging to this polytope can be fully specified by $6$ singlepartite, $12$ bipartite and $8$ tripartite expectations,
\begin{align}
P(abc|A_x B_y C_z)= & \frac{1}{8}[1+(-1)^a\braket{A_x}+(-1)^b\braket{B_y}+(-1)^c\braket{C_z}
+(-1)^{a\oplus b}\braket{A_xB_y}+(-1)^{a\oplus c}\braket{A_xC_z}+(-1)^{b\oplus c}\braket{B_yC_z} \nonumber \\
& +(-1)^{a\oplus b\oplus c}\braket{A_xB_yC_z}],
\end{align}
where $\braket{A_x}=\sum_a (-1)^a P(a|A_x)$, $\braket{A_xB_y}=\sum_{a,b}(-1)^{a\oplus b}P(ab|A_x B_y)$ and $\braket{A_xB_yC_z}=\sum_{a,b,c}(-1)^{a\oplus b \oplus c}P(abc|A_x B_y C_z)$, $\oplus$ denotes modulo sum $2$.
The set of boxes that can be simulated by a fully LHV model are of the form,
\begin{align}
P(abc|A_x B_y C_z)=\sum^{d_\lambda-1}_{\lambda=0} p_\lambda P_\lambda(a|A_x)P_\lambda(b|B_y)P_\lambda(c|C_z), \label{LHV}
\end{align}
which form a fully local polytope \cite{ns2, ns3} denoted by $\mathcal{L}$. 
Here $\lambda$ denotes shared classical randomness/local hidden variable (LHV) which occurs with probability $p_\lambda$. 
For a given fully local box, the form (\ref{LHV}) determines a 
classical simulation protocol with dimension $d_\lambda$ \cite{sl1}.
The extremal boxes of $\mathcal{L}$ are $64$ fully  local vertices which are fully deterministic boxes given by,
\begin{equation}
P^{\alpha\beta\gamma\epsilon\zeta\eta}_D(abc|A_x B_y C_z)=\left\{
\begin{array}{lr}
1, & a=\alpha x\oplus \beta\\
   & b=\gamma y\oplus \epsilon \\
   & c=\zeta  z\oplus \eta\\
0 , & \text{otherwise}.\\
\end{array}
\right.  \label{DB} 
\end{equation}
Here, $\alpha,\beta,\gamma, \epsilon, \zeta, \eta \in\{0,1\}$. The above boxes can be written as the 
product of deterministic distributions corresponding to Alice and Bob-Charlie,  i.e.,
$P^{\alpha\beta\gamma\epsilon\zeta\eta}_D(abc|A_x B_y C_z)=P^{\alpha\beta}_D(a|A_x)P^{\gamma\epsilon\zeta\eta}_D(bc|B_y C_z)$,
where 

\begin{equation}
P^{\alpha\beta}_D(a|A_x)=\left\{
\begin{array}{lr}
1, & a=\alpha x\oplus \beta\\
0 , & \text{otherwise}\\
\end{array}
\right.   
\label{dba}
\end{equation}
and
\begin{equation}
P^{\gamma\epsilon\zeta\eta}_D(bc|B_y C_z)=\left\{
\begin{array}{lr}
1, & b=\gamma y\oplus \epsilon \\
   & c=\zeta  z\oplus \eta\\
0 , & \text{otherwise},\\
\end{array}
\right.  \label{DB} \end{equation}
which can also be written as the product of deterministic distributions corresponding to Bob and Charlie, 
i.e., $P^{\gamma\epsilon\zeta\eta}_D(bc|B_y C_z)=P^{\gamma\epsilon}_D(b|B_y)P^{\zeta\eta}_D(c|C_z)$, where
\begin{equation}
P^{\gamma\epsilon}_D(b|B_y)=\left\{
\begin{array}{lr}
1, & b=\gamma y\oplus \epsilon\\
0 , & \text{otherwise}\\
\end{array}
\right.   
\label{db1}
\end{equation}
and 
\begin{equation}
P^{\zeta\eta}_D(c|C_z)=\left\{
\begin{array}{lr}
1, & c=\zeta  z\oplus \eta\\
0 , & \text{otherwise.}\\
\end{array}
\right.  
\label{dc1} 
\end{equation}
Hence, one can write, $P^{\alpha\beta\gamma\epsilon\zeta\eta}_D(abc|A_x B_y C_z)=P^{\alpha\beta}_D(a|A_x)P^{\gamma\epsilon}_D(b|B_y)P^{\zeta\eta}_D(c|C_z)$.\\

The set of local boxes and quantum boxes satisfy $\mathcal{L} \subset Q \subset \mathcal{N}$.
Boxes lying outside $\mathcal{L}$ are called nonlocal boxes and they cannot be written as a convex mixture of the local deterministic boxes alone.

Nonlocal boxes can be classified into two categories:\\
i) genuinely three-way nonlocal and\\
ii) two-way local boxes.\\
A nonlocal box is genuinely three-way nonlocal \emph{if and only if} (iff) it cannot be written in the two-way local form \cite{ns4} given by,
\begin{align}
P(abc|A_x B_y C_z)&=p_1\sum_\lambda r_\lambda P_\lambda^{A|BC} +p_2\sum_\lambda s_\lambda P_\lambda^{B|AC} + p_3 \sum_\lambda t_\lambda P_\lambda^{C|AB} , \label{HLNL}
\end{align}
where, $P_\lambda^{A|BC}= P_\lambda(a|A_x)\, P_\lambda(bc|B_y C_z)$, and, $P_\lambda^{B|AC}$ and $P_\lambda^{C|AB}$  are  similarly  defined; $\sum_\lambda r_\lambda$ = $\sum_\lambda s_\lambda$ = $\sum_\lambda t_\lambda$ = $1$; $p_1 + p_2 + p_3 = 1$.
Each bipartite distribution in this decomposition can have arbitrary nonlocality consistent with the NS principle. Following \cite{BL}, we define a tripartite correlation $P(abc|A_x B_y C_z)$ as no-signalling bilocal (NSBL) across the bipartition $(A|BC)$ iff it has the following decomposition:
\begin{equation}
P(abc|A_x B_y C_z)= \sum_\lambda r_\lambda P_\lambda(a|A_x)\,P_\lambda(bc|B_yC_z).
\end{equation} 
Hence, a genuinely three-way nonlocal box is not NSBL across any possible bipartition. The set of boxes that admit a decomposition as in Eq. (\ref{HLNL}) again forms a convex polytope, which is called two-way local polytope denoted by $\mathcal{L}_2$.
The extremal boxes of this polytope are the $64$ local vertices and $48$ two-way local vertices.
There are $16$ two-way local vertices in which a PR-box \cite{pr} is shared between $A$ and $B$,
\begin{align}
&P^{\alpha\beta\gamma\epsilon}_{12}(abc|A_x B_y C_z)
=\left\{
\begin{array}{lr}
\frac{1}{2}, & a\oplus b=x\cdot y \oplus \alpha x\oplus \beta y \oplus \gamma \quad \& \quad c=\gamma z \oplus \epsilon\\
0 , & \text{otherwise},\\
\end{array}
\right. \label{PR}
\end{align}
the other $32$ two-way local vertices, $P^{\alpha\beta\gamma\epsilon}_{13}$ and $P^{\alpha\beta\gamma\epsilon}_{23}$,
in which a PR-box is shared by $AC$ and $BC$, respectively, are similarly defined. The extremal boxes in Eq. (\ref{PR}) can be written in the factorized form,
$P^{\alpha\beta\gamma\epsilon}_{12}(abc|A_x B_y C_z)=P^{\alpha\beta\gamma}_{PR}(ab|A_x B_y)P^{\gamma\epsilon}_D(c|C_z)$, where
$P^{\alpha\beta\gamma}_{PR}(ab|A_x B_y)$ are the $8$ PR-boxes given by,
\[
    P_{PR}^{\alpha \beta \gamma} (ab|A_x B_y) = 
\begin{dcases}
    \frac{1}{2},& \text{if } a \oplus b = x.y \oplus \alpha x \oplus \beta y \oplus \gamma \\
    0,              & \text{otherwise},
\end{dcases}
\] 
and 
\begin{equation}
P^{\zeta\eta}_D(c|C_z)=\left\{
\begin{array}{lr}
1, & c=\gamma z \oplus \epsilon\\
0 , & \text{otherwise.}\\
\end{array}
\right.  
\end{equation}
Though in the above the two-way local boxes are defined only for nonlocal boxes, fully local boxes are also two-way local.
The set of two-way local boxes satisfy, $\mathcal{L} \subset \mathcal{L}_2 \subset \mathcal{N}$.
A genuinely three-way nonlocal box cannot be written as a convex mixture of the extremal boxes of $\mathcal{L}_{2}$ alone and violates
a facet inequality of $\mathcal{L}_2$ given in Ref. \cite{ns4}.

The Svetlichny inequalities \cite{SI} which are given by
\begin{equation}
\mathcal{S}_{\alpha\beta\gamma\epsilon}
=\sum_{xyz}(-1)^{x\cdot y \oplus x\cdot z \oplus y\cdot z \oplus \alpha x\oplus \beta y \oplus \gamma z \oplus \epsilon}\braket{A_xB_yC_z}\le4, \label{SI}
\end{equation}
are one of the classes of facet inequalities of the two-way local polytope.
The violation of a Svetlichny inequality implies one of the forms of genuine nonlocality.
The following extremal three-way nonlocal boxes:
\begin{align}
&P^{\alpha\beta\gamma\epsilon}_{\rm Sv}(abc|A_x B_y C_z) \nonumber \\
&=\left\{
\begin{array}{lr}
\frac{1}{4}, & \!a\!\oplus \!b\!\oplus \!c\!
=\!x\cdot y \!\oplus \!x\cdot z\! \oplus \!y\cdot z \!\oplus \!\alpha x\!\oplus\! \beta y\! \oplus\! \gamma z \!\oplus\! \epsilon\\
0 , & \text{otherwise},\\
\end{array}
\right. \label{NLV}
\end{align}
which violate a Svetlichny inequality to its algebraic maximum are called Svetlichny boxes. Boxes that violate a Svetlichny inequality and do not violate any Svetlichny inequality are called Svetlichny nonlocal and
Svetlichny-local boxes, respectively.
Mermin inequalities \cite{mermin} are one of the classes 
of facet inequalities of the fully local polytope \cite{ns5, ns6}.
One of the Mermin inequalities is given by,
\begin{equation}
\braket{A_0B_0C_0}-\braket{A_0B_1C_1}-\braket{A_1B_0C_1}-\braket{A_1B_1C_0}\le2, \label{MI}
\end{equation}
and the other $15$ Mermin inequalities can be obtained from the above inequality
by local reversible operations (LRO), which are analogous to local unitary operations in quantum theory
and include local relabeling of the inputs and outputs (conditionally on the input).
Mermin inequalities detect certain nonlocal boxes which are two-way local.
Quantum correlations that violate a Mermin inequality to its algebraic maximum 
demonstrate  Greenberger--Horne--Zeilinger (GHZ) paradox \cite{GHZ} and are called Mermin boxes.

If a Svetlichny nonlocal box is decomposed in the context of NS polytope, then it necessarily has a Svetlichny-box fraction in the decomposition.  In \cite{mp4}, the author studied a canonical decomposition 
for the Svetlichny nonlocal boxes.
According to this decomposition, a given Svetlichny nonlocal box 
is written as a convex mixture of an irreducible Svetlichny-box and a Svetlichny-local box $P_{SvL}$ 
 without having the Svetlichny box $P^{\alpha\beta\gamma\epsilon}_{Sv}$ fraction excessively, i.e.,
\begin{equation}
P=p_{Sv}P^{\alpha\beta\gamma\epsilon}_{Sv}+(1-p_{Sv})P_{SvL},
\end{equation}
where $p_{Sv}$ is called  Svetlichny strength  which signifies the maximal Svetlichny-box fraction of a Svetlichny-nonlocal box.
Similarly, in \cite{mp4}, the author defined Mermin strength as the Mermin-box fraction of a Mermin-nonlocal box (which violates a Mermin inequality) in its canonical decomposition which is a convex combinations of one dominant Mermin-box  and one Mermin-local box (which does not violate any Mermin inequality) without having the Mermin box fraction excessively. Svetlichny strength and/or Mermin strength can also be nonzero for certain local correlations \cite{mp4}.

\subsection{Definitions of bipartite and genuine tripartite quantum steering}

\textbf{Bipartite quantum steering:} Let us consider a steering scenario where two spatially separated parties, say Alice and Bob, share an unknown quantum system $\rho_{AB}\in \mathcal{B}(\mathcal{H}_A \otimes \mathcal{H}_B)$ with the Hilbert-space dimension of Bob's 
subsystem is known and  Alice 
performs a set of black-box measurements to demonstrate steerability to Bob's conditional states prepared by him. 
Such a measurement scenario is called one sided device independent since
Alice's measurement operators $\{M_{a|A_x}\}_{a,A_x}$ are unknown. Let  $\{\sigma_{a|A_x}\}_{a,A_x}$ denote the set of unnormalized conditional states on Bob's side prepared by Alice's measurements and is called an  assemblage. Each element
in the assemblage is given by $\sigma_{a|A_x}=P(a|A_x)\rho_{a|A_x}$,  where $P(a|A_x)$ is the conditional probability of getting the outcome $a$ when Alice performs the measurement $A_x$;
$\rho_{a|A_x}$ is the normalized conditional state on Bob's side.
Quantum theory predicts that all valid assemblages should satisfy the following criteria:
\begin{equation}
\sigma_{a|A_x}= Tr_A ( M_{a|A_x} \otimes \openone \rho_{AB}) \hspace{0.5cm} \forall \sigma_{a|A_x} \in \{\sigma_{a|A_x}\}_{a,A_x}.
\end{equation}

In the above scenario, Alice demonstrates steerability to Bob
\textit{iff} the assemblage does not have a local hidden state (LHS) model, i.e., if for all $a$, $A_x$, there
is no decomposition of $\sigma_{a|A_x}$ in the form,
\begin{equation}
\sigma_{a|A_x}=\sum_\lambda r_{\lambda} P_{\lambda}(a|A_x) \rho^\lambda_B,
\end{equation}
where $\lambda$ denotes classical random variable which occurs with probability 
$r_{\lambda}$; $\sum_{\lambda} r_{\lambda} = 1$; $\rho^\lambda_B$
are called local hidden states which satisfy $\rho^\lambda_B\ge0$ and
Tr$\rho^\lambda_B=1$.

Suppose Bob performs a set of projective measurements $\{\Pi_{b|B_y}\}_{b,B_y}$ on $\{\sigma_{a|A_x}\}_{a,A_x}$ producing measurement correlations $P_{\rho_{AB}}(ab|A_x B_y)$, where $P_{\rho_{AB}}(ab|A_x B_y)$ = $Tr ( \Pi_{b|B_y} \sigma_{a|A_x} )$. The correlations $P_{\rho_{AB}}(ab|A_x B_y)$ detects steerability from Alice to Bob \textit{iff} it does not have a decomposition as follows \cite{steer, steer2}: 
\begin{equation}
P_{\rho_{AB}}(ab|A_x B_y)= \sum_\lambda r_{\lambda} P_{\lambda}(a|A_x) P(b|B_y, \rho^\lambda_B) \hspace{0.3cm} \forall a,A_x,b,B_y; \label{LHV-LHS}
\end{equation}
where, $\sum_{\lambda} r_{\lambda} = 1$, $P_{\lambda}(a|A_x)$ denotes an arbitrary probability distribution (deterministic/non-deterministic boxes) arising from local hidden variable (LHV) $\lambda$ ($\lambda$ occurs with probability $r_{\lambda}$) and $P(b|B_y, \rho^{\lambda}_B) $ = Tr$(\Pi_{b|B_y} \rho^{\lambda}_B)$ denotes the quantum probability of outcome $b$ when measurement $B_y$ is performed on local hidden state (LHS) $\rho^{\lambda}_B$. \\

\textbf{Genuine tripartite quantum steering:} Before we define the notion of genuine tripartite quantum steering as introduced in 
Ref. \cite{st10}, we define bi-unsteerability for the tripartite  one sided device independent scenario where one of the parties performs black-box measurements
and the other two parties perform trusted measurements. 
Suppose, three spatially separated parties (say, Alice, Bob and Charlie) share a quantum mechanical system $\rho_{ABC} \in \mathcal{B}( \mathcal{H}_A \otimes \mathcal{H}_B \otimes \mathcal{H}_C)$. Let us assume that the tripartite correlations $P_{\rho_{ABC}}(abc|A_x B_ y C_z)$ is produced when Alice performs a set of black-box measurements $\{M_{a|A_x}\}_{a,A_x}$; Bob and Charlie perform quantum projective measurements $\{\Pi_{b|B_y}\}_{b,B_y}$ and $\{\Pi_{c|C_z}\}_{c,C_z}$ respectively. The tripartite correlations $P_{\rho_{ABC}}(abc|A_x B_y C_z)$ is called bi-unsteerable across the bipartite cut 
$A-BC$ if it admits a decomposition of the form: 

%sets of observables in the Hilbert space of Alice, Bob and Charlie's system are denoted by $\mathcal{D}_{\alpha}$, $\mathcal{D}_{\beta}$ and $\mathcal{D}_{\gamma}$ respectively. $A_x$ denotes an element of $\mathcal{D}_{\alpha}$, and the set of outcomes are labeled by $a \in \mathcal{L}(A)$. Similarly, $B_y$ denotes an element of $\mathcal{D}_{\beta}$, and the set of outcomes are labeled by $b \in \mathcal{L}(B)$; $C_z$ denotes an element of $\mathcal{D}_{\gamma}$, and the set of outcomes are labeled by $c \in \mathcal{L}(C)$. The joint state $\rho_{ABC}$ of the system shared between Alice, Bob and Charlie is bi-steerable from Alice to Bob-Charlie iff for all $a \in \mathcal{L}(A)$, $b \in \mathcal{L}(B)$, $c \in \mathcal{L}(C)$, $A_x \in \mathcal{D}_{\alpha}$, $B_y \in \mathcal{D}_{\beta}$, $C_z \in \mathcal{D}_{\gamma}$, the joint probability distributions \textit{cannot} be written in the form:

\begin{equation}
P_{\rho_{ABC}}(abc|A_x B_ y C_z) = \sum_{\lambda} r_{\lambda} P_{\lambda} (a|A_x) P(b c |B_y, C_z,  \rho^{\lambda}_{BC}),
\label{bisteer}
\end{equation}
with $\sum_{\lambda} r_{\lambda} = 1$. Here, $P_{\lambda} (a|A_x)$ denotes an arbitrary probability distribution (deterministic/non-deterministic boxes) arising from local hidden variable (LHV) $\lambda$ ($\lambda$ occurs with probability $r_{\lambda}$) and $P(b c |B_y, C_z,  \rho^{\lambda}_{BC}) $ = Tr$(\Pi_{b|B_y} \otimes \Pi_{c|C_z} \rho^{\lambda}_{BC})$ denotes the quantum probability of obtaining the outcomes $b$ and $c$, when measurements $B_y$ and $C_z$ are performed by Bob and Charlie, respectively, on the bipartite local hidden state (LHS) $\rho^{\lambda}_{BC}$ shared between Bob and Charlie. The quantum probability distribution $P(b c |B_y, C_z,  \rho^{\lambda}_{BC}) $ can demonstrate quantum nonlocality, or EPR-steering (from Bob to Charlie, or from Charlie to Bob, or both), or locality, or unsteerability. 
Similarly one can define bi-unsteerability for the the tripartite correlations $P_{\rho_{ABC}}(abc|A_x B_y C_z)$ across the other two bipartite
cuts in the respective one sided device independent scenarios.
A tripartite correlation $P_{\rho_{ABC}}(abc|A_x B_y C_z)$ 
which does not have a bi-unsteerable form may have genuine tripartite steerability.  Note that the bi-unsteerable correlations form a subset of the two-way local correlations, as the bipartite distributions in the two-way local correlations are NS box and the bipartite distributions in the bi-unsteerable correlations are quantum correlations. We define bi-unsteerability motivated by the the definition of bilocal correlation introduced in the context of genuine multipartite nonlocality by Gallego et. al. \cite{BL}.

The tripartite correlations $P_{\rho_{ABC}}(abc|A_x B_y C_z)$ detects genuine tripartite steerability \textit{iff} it cannot be written as a convex combination of bi-unsteerable correlations in all three possible  bipartitions. In other words, the tripartite correlations $P_{\rho_{ABC}}(abc|A_x B_y C_z)$ detects genuine steering \textit{iff} it does not have a decomposition as follows \cite{st10, stm7}:
\begin{align}
P_{\rho_{ABC}}(abc|A_x B_ y C_z) = &p_1 \sum_{\lambda} r_{\lambda} P_{\lambda} (a|A_x) P(b c |B_y, C_z,  \rho^{\lambda}_{BC}) + p_2 \sum_{\lambda} s_{\lambda} P_{\lambda} (b|B_y) P(a c |A_x, C_z,  \rho^{\lambda}_{AC}) \nonumber\\
&+ p_3 \sum_{\lambda} t_{\lambda} P_{\lambda} (c|C_z) P(a b |A_x, B_y,  \rho^{\lambda}_{AB}),
\label{genuinesteering}
\end{align}
where $p_1 + p_2 + p_3 =1$, $\sum_{\lambda} r_{\lambda} = 1$, $\sum_{\lambda} s_{\lambda} = 1$, $\sum_{\lambda} t_{\lambda} = 1$. The single-partite and bipartite distributions are defined in a similar way as mentioned earlier.

Note that in each term in Eq.(\ref{genuinesteering}), the single-partite terms are arbitrary and the bipartite terms are restricted to be quantum. This is due to the fact that for each bi-unsteerable term in Eq.(\ref{genuinesteering}) we have considered the steering scenario where one party steers the other two parties' joint state (one-to-two steering scenario or one sided device independent scenario) \cite{stm7}. Similarly, genuine tripartite steering can also be defined where each bi-unsteerable term in the convex combination is defined in the steering scenario where two parties jointly steer the third party's state (two-to-one steering scenario or two sided device independent scenario) \cite{st10}. 

The above definition of genuine steering has been demonstrated experimentally in \cite{stm4, stm42}.

\section{Definition of genuine super-bi-unsteerability}

For a given bipartite or n-partite box, let $d_{\lambda}$ denotes the minimal dimension of the shared classical randomness. Before we define super-bi-unsteerability for bi-unsteerable tripartite boxes, let us recapitulate the notion of super-unsteerability \cite{sus} for unsteerable bipartite boxes.\\

\textbf{Definition 1:} \textit{Suppose two spatially separated party (say, Alice and Bob) share a bipartite quantum state $\rho_{AB}$ in $\mathbb{C}^{d^A} \otimes \mathbb{C}^{d^B}$ producing a correlation box $P(a b|A_x B_y)$ which is unsteerable from Alice to Bob. Then, super-unsteerability holds iff there is no decomposition of the form:
\begin{equation}
\label{sup}
P(a b|A_x B_y) = \sum_{\lambda = 0}^{d_{\lambda} -1} r_{\lambda} P_{\lambda}(a|A_x) P (b|B_y,  \rho^{\lambda}_B),
\end{equation}
where $d_{\lambda} \leq d^A$. Here, $P_{\lambda}(a|A_x)$ denotes an arbitrary probability distribution (deterministic/non-deterministic boxes) arising from local hidden variable (LHV) $\lambda$ and $P (b|B_y,  \rho_{\lambda}^B)$ are the quantum probability of obtaining the outcome $b$, when measurement $B_y$ is performed by Bob on LHS $\rho^{\lambda}_{B}$ in $\mathbb{C}^{d^B}$; $\sum_{\lambda = 0}^{d_{\lambda} -1} r_{\lambda} = 1$.}\\

We now define super-bi-unsteerability for the bi-unsteerable tripartite boxes.\\

\textbf{Definition 2:} \textit{Suppose three spatially separated party (say, Alice, Bob and Charlie) share a tripartite quantum state $\rho'_{ABC}$ in $\mathbb{C}^{d^A} \otimes \mathbb{C}^{d^B} \otimes \mathbb{C}^{d^C}$ producing a correlation box $P(a b c|A_x B_y C_z)$ which is bi-unsteerable from Alice to Bob-Charlie. Then super-bi-unsteerability from Alice to Bob-Charlie holds iff there is no decomposition of the form:
\begin{equation}
\label{bisup}
P(a b c|A_x B_y C_z) = \sum_{\lambda = 0}^{d_{\lambda} -1} r_{\lambda} P_{\lambda}(a|A_x) P (b c|B_y C_z, \rho^{\lambda}_{BC}),
\end{equation}
where $d_{\lambda} \leq d^A$.
Here, $P_{\lambda}(a|A_x)$ denotes an arbitrary probability distribution (deterministic/non-deterministic boxes) arising from local hidden variable (LHV) $\lambda$ and $P (b c|B_y C_z, \rho^{\lambda}_{BC})$ are the quantum probability of obtaining the outcomes $b$ and $c$, when measurements $B_y$ and $C_z$ are performed by Bob and Charlie, respectively, on the bipartite LHS $\rho^{\lambda}_{BC}$ in $\mathbb{C}^{d^B} \otimes \mathbb{C}^{d^C}$; $\sum_{\lambda = 0}^{d_{\lambda} -1} r_{\lambda} = 1$. $P (b c|B_y C_z, \rho^{\lambda}_{BC})$  may demonstrate quantum nonlocality or EPR-steering.}\\

Super-bi-unsteerability across other bipartitions can be defined similarly.

Quantumness of certain bipartite unsteerable correlation has been operationally characterized by the notion of super-unsteerability \cite{sus} and it has been demonstrated that bipartite quantum discord \cite{disc, disc2, disc3} is necessary for demonstrating bipartite super-unsteerability \cite{sus}. In the tripartite scenario, genuine tripartite quantum discord was defined in order to quantify the genuine quantumness of  tripartite quantum states \cite{mp1}. In \cite{mp2} Zhao et. al. defined genuine tripartite quantum discord as the minimum bipartite discord over all possible bipartitions. Hence, any tripartite state has non-zero genuine tripartite discord iff it has non-zero bipartite discord across all possible bipartitions. Motivated by these facts, we define genuine super-bi-unsteerability of tripartite correlations as follows. \\

\textbf{Definition 3:} \textit{ A tripartite bi-unsteerable correlation is said to be genuinely super-bi-unsteerable \textit{iff} it is super-bi-unsteerable across all possible bipartitions (i. e., from Alice to Bob-Charlie, from Bob to Alice-Charlie, and from Charlie to Alice-Bob).}\\

In the present study, as mentioned earlier, we have restricted ourselves to one-to-two steering scenario or one sided device independent scenario. That is why the single-partite term in Eq.(\ref{bisup}) is arbitrary and the bipartite term in Eq.(\ref{bisup}) is restricted to be quantum. In a similar way one can define super-bi-unsteerability in two-to-one steering scenario or two sided device independent scenario. In this case, the bipartite distributions will be an arbitrary NS box and and single-partite distribution will be quantum.

Another important point to be stressed here is that super-bi-unsteerability across a particular bipartition is not a genuine multipartite property. This definition is not invariant under permutation of parties. On the other hand, genuine super-bi-unsteerability is not defined across a particular bipartition and, hence, is invariant under permutation of parties.

In the following Section, we are going to study some specific examples of genuine super-bi-unsteerability  in one-to-two steering scenario or one sided device independent scenario.

\section{Specific examples of genuine super-bi-unsteerability}

We consider quantum correlations that belong to the noisy Mermin family defined as

\begin{equation}
P_{MF}^{V} (abc|A_x B_ y C_z) = \frac{1 + (-1)^{a \oplus b \oplus c \oplus  xy \oplus yz \oplus xz} \delta_{x \oplus y \oplus 1,z} V}{8}  ,
\label{MFO}
\end{equation}
where $0 < V \leq 1$. The above box is two-way local, but not fully local for $V > \frac{1}{2}$ as it violates the Mermin inequality (given in Eq. (\ref{MI})) in this range, and for $V \leq \frac{1}{2}$, it is fully local as in this range the correlation does not violate any Bell inequality. Note that for any $V> 0$, the quantum simulation of the Mermin family by using a $2 \otimes 2 \otimes 2$ quantum state necessarily requires genuine quantumn discord \cite{mp1, mp2} in the state. Because, the Mermin family has nonzero Mermin strength for any $V > 0$ \cite{mp4, mp10}. We now give example of simulating the noisy Mermin family by using a quantum state which has quantumness. Consider, the three spatially separated parties (say, Alice, Bob and Charlie) share the following $2 \otimes 2 \otimes 2$ GHZ state:
\begin{equation}
\label{GHZ}
\rho_1 = V | GHZ \rangle \langle GHZ | + (1-V) \frac{\mathbb{I}_2}{2} \otimes \frac{\mathbb{I}_2}{2} \otimes \frac{\mathbb{I}_2}{2},
\end{equation}
where $| GHZ \rangle = \frac{1}{\sqrt{2}} (|000 \rangle + |111\rangle)$; $0 < V \leq 1$; $|0\rangle$ and $|1\rangle$ are the eigenstates of operator $\sigma_z$ corresponding to eigenvalues $+1$ and $-1$ respectively; $\mathbb{I}_2$ is the $2 \otimes 2$ identity matrix. Then the noisy Mermin family can be reproduced if Alice, Bob and Charlie perform projective qubit measurement corresponding to the operators: $A_0 = \sigma_y$, $A_1 = - \sigma_x$; $B_0 = \sigma_y$, $B_1 = - \sigma_x$; $C_0 = \sigma_y$, $C_1 = - \sigma_x$ respectively. Hence, noisy Mermin family can be simulated with $2 \otimes 2 \otimes 2$ quantum states.

\subsection{Simulating noisy Mermin family with LHV at Alice's side and LHS at Bob-Charlie's side}
The correlation belonging to noisy Mermin family can be written as
\begin{equation}
\label{lhvlhs}
P_{MF}^{V} (abc|A_x B_ y C_z) = \sum_{\lambda=0}^{3} r_{\lambda} P_{\lambda} (a|A_x) P(b c |B_y, C_z,  \rho^{\lambda}_{BC}),
\end{equation}
where $r_0$ = $r_1$ = $r_2$ = $r_3$ = $\frac{1}{4}$, and \\
$P_{0} (a|A_x)$ = $P_D^{00}$, $P_{1} (a|A_x)$ = $P_D^{01}$, $P_{2} (a|A_x)$ = $P_D^{10}$, $P_{3} (a|A_x)$ = $P_D^{11}$.\\
where, \begin{equation}
P_D^{\alpha\beta}(a|A_x)=\left\{
\begin{array}{lr}
1, & a=\alpha x\oplus \beta\\
0 , & \text{otherwise}.\\
\end{array}
\right. 
\label{}
\end{equation}

Now,
 \begin{equation}
 P(b c |B_y, C_z,  \rho^{0}_{BC}) = \bordermatrix{
\frac{bc}{yz} & 00 & 01 & 10 & 11 \cr
00 & \frac{1+V}{4} & \frac{1-V}{4} & \frac{1-V}{4} & \frac{1+V}{4} \cr
01 & \frac{1+V}{4} & \frac{1-V}{4} & \frac{1-V}{4} & \frac{1+V}{4} \cr
10 & \frac{1+V}{4} & \frac{1-V}{4} & \frac{1-V}{4} & \frac{1+V}{4} \cr
11 & \frac{1-V}{4} & \frac{1+V}{4} & \frac{1+V}{4} & \frac{1-V}{4} } ,
\end{equation}
where each row and column corresponds to a fixed measurement settings $(yz)$ and a fixed outcome $(bc)$ respectively. Throughout the paper we will follow the same convention.

This joint probability distribution at Bob and Charlie's side can be reproduced by performing the projective qubit measurements of the observables corresponding to the operators $B_0 = \sigma_y$, $B_1 = - \sigma_x$; and $C_0 = \sigma_y$, $C_1 = - \sigma_x$ on the state given by
\begin{equation}
\label{state1}
\rho^0_{BC} =  | \psi_0 \rangle \langle \psi_0 | ,
\end{equation}
where, $| \psi_0 \rangle = \cos \theta |00 \rangle - \dfrac{1+i}{\sqrt{2}} \sin \theta |11 \rangle$ ($0 \leq \theta \leq \frac{\pi}{4}$) with $\sin 2 \theta = \sqrt{2} V$; $|0\rangle$ and $|1\rangle$ are the eigenstates of $\sigma_z$ corresponding to the eigenvalues $+1$ and $-1$ respectively.

\begin{center}
$P(b c |B_y, C_z,  \rho^{1}_{BC}) = \begin{pmatrix}
\frac{1-V}{4} && \frac{1+V}{4} && \frac{1+V}{4} && \frac{1-V}{4}\\
\frac{1-V}{4} && \frac{1+V}{4} && \frac{1+V}{4} && \frac{1-V}{4}\\
\frac{1-V}{4} && \frac{1+V}{4} && \frac{1+V}{4} && \frac{1-V}{4}\\
\frac{1+V}{4} && \frac{1-V}{4} && \frac{1-V}{4} && \frac{1+V}{4}\\
\end{pmatrix}, $ 
\end{center} 

This joint probability distribution at Bob and Charlie's side can be reproduced by performing the projective qubit measurements of the observables corresponding to the operators $B_0 = \sigma_y$, $B_1 = - \sigma_x$; and $C_0 = \sigma_y$, $C_1 = - \sigma_x$ on the state given by
\begin{equation}
\label{state2}
\rho^1_{BC} =  | \psi_1 \rangle \langle \psi_1 | ,
\end{equation}
where, $| \psi_1 \rangle =   \cos \theta |00 \rangle + \dfrac{1+i}{\sqrt{2}} \sin \theta |11 \rangle$ ($0 \leq \theta \leq \frac{\pi}{4}$) with $\sin 2 \theta = \sqrt{2} V$.

\begin{center}
$P(b c |B_y, C_z,  \rho^{2}_{BC}) = \begin{pmatrix}
\frac{1-V}{4} && \frac{1+V}{4} && \frac{1+V}{4} && \frac{1-V}{4}\\
\frac{1+V}{4} && \frac{1-V}{4} && \frac{1-V}{4} && \frac{1+V}{4}\\
\frac{1+V}{4} && \frac{1-V}{4} && \frac{1-V}{4} && \frac{1+V}{4}\\
\frac{1+V}{4} && \frac{1-V}{4} && \frac{1-V}{4} && \frac{1+V}{4}\\
\end{pmatrix} ,$ 
\end{center}

This joint probability distribution at Bob and Charlie's side can be reproduced by performing the projective qubit measurements of the observables corresponding to the operators $B_0 = \sigma_y$, $B_1 = - \sigma_x$; and $C_0 = \sigma_y$, $C_1 = - \sigma_x$ on the state given by
\begin{equation}
\label{state3}
\rho^2_{BC} =  | \psi_2 \rangle \langle \psi_2 | ,
\end{equation}
where, $| \psi_2 \rangle =  \cos \theta |00 \rangle + \dfrac{1-i}{\sqrt{2}} \sin \theta |11 \rangle$ ($0 \leq \theta \leq \frac{\pi}{4}$) with $\sin 2 \theta = \sqrt{2} V$.

\begin{center}
$P(b c |B_y, C_z,  \rho^{3}_{BC}) = \begin{pmatrix}
\frac{1+V}{4} && \frac{1-V}{4} && \frac{1-V}{4} && \frac{1+V}{4}\\
\frac{1-V}{4} && \frac{1+V}{4} && \frac{1+V}{4} && \frac{1-V}{4}\\
\frac{1-V}{4} && \frac{1+V}{4} && \frac{1+V}{4} && \frac{1-V}{4}\\
\frac{1-V}{4} && \frac{1+V}{4} && \frac{1+V}{4} && \frac{1-V}{4}\\
\end{pmatrix} ,$ 
\end{center}

This joint probability distribution at Bob and Charlie's side can be reproduced by performing the projective qubit measurements of the observables corresponding to the operators $B_0 = \sigma_y$, $B_1 = - \sigma_x$; and $C_0 = \sigma_y$, $C_1 = - \sigma_x$ on the state given by
\begin{equation}
\label{state4}
\rho^3_{BC} =  | \psi_3 \rangle \langle \psi_3 | ,
\end{equation}
where, $| \psi_3 \rangle = \cos \theta |00 \rangle - \dfrac{1-i}{\sqrt{2}} \sin \theta |11 \rangle$ ($0 \leq \theta \leq \frac{\pi}{4}$) with $\sin 2 \theta = \sqrt{2} V$.

Now, $|\sin 2 \theta | \leq 1$ (as $0 \leq \theta \leq \frac{\pi}{4}$), which implies that $V \leq \frac{1}{\sqrt{2}}$. Hence, one can state that the noisy Mermin family can be expressed with a LHV-LHS decomposition (\ref{lhvlhs})  from Alice to Bob-Charlie in one sided device independent scenario having hidden variables of  dimension $4$ in the range $0 < V \leq \frac{1}{\sqrt{2}}$. The noisy Mermin family for $V \leq \frac{1}{\sqrt{2}}$, therefore, is bi-unsteerable in the bipartition $A-BC$ in one sided device independent scenario. Each joint probability distribution at Bob-Charlie's side $P(b c |B_y, C_z,  \rho^{\lambda}_{BC})$ ($\lambda =0,1,2,3$) produced from the LHS demonstrates EPR-steering when $\frac{1}{2} < V \leq \frac{1}{\sqrt{2}}$ (if the two measurement settings of the party which is being steered are mutually unbiased), because in this range, each of the
$P(b c |B_y, C_z,  \rho^{\lambda}_{BC})$ violates the analogous Clauser-Horne-Shimony-Holt inequality for steering \cite{stns}. Each joint probability distribution at Bob-Charlie's side $P(b c |B_y, C_z,  \rho^{\lambda}_{BC})$ produced from the LHS demonstrates super-unsteerability when $0 < V \leq \frac{1}{2}$ (for detailed calculations, see the Appendix \ref{a1}). Since noisy Mermin box is invariant under permutations of parties, it can be stated that the noisy Mermin family for $V \leq \frac{1}{\sqrt{2}}$ is bi-unsteerable in the bipartitions $B-AC$ and $C-AB$ in one sided device independent scenario.

Hence, the decomposition (\ref{lhvlhs}) represents a LHV-LHS decomposition of the bi-unsteerable (from Alice to Bob-Charlie) noisy Mermin box with different deterministic distributions at Alice's side for $0 < V \leq \frac{1}{\sqrt{2}}$  in one sided device independent scenario.\\

\textbf{Theorem 1.} \textit{The LHV-LHS decomposition of bi-unsteerable
noisy Mermin box  from Alice to Bob-Charlie in one sided device independent scenario cannot be realized with hidden variables having dimension $3$ for $V > \frac{1}{\sqrt{5}}$}\\

\textit{Proof.} Let us try to generate a LHV-LHS decomposition of the bi-unsteerable noisy Mermin family  from Alice to Bob-Charlie in one sided device independent scenario with hidden variables having dimension $3$ and with different deterministic distributions at Alice's side. Before proceeding, we want to mention that in case of noisy Mermin family, all the marginal probability distributions of Alice, Bob and Charlie are maximally mixed:
\begin{equation}
\label{mar}
P(a|A_x) = P(b|B_y) = P(c|C_z) = \frac{1}{2} \forall a,b,c,x,y,z.
\end{equation} 
Consider that the noisy Mermin family can be decomposed in the following way:
\begin{equation}
P_{MF}^{V} (abc|A_x B_ y C_z) = \sum_{\lambda=0}^{2} r_{\lambda} P_{\lambda} (a|A_x) P(b c |B_y, C_z,  \rho^{\lambda}_{BC}).
\end{equation}
Here, $r_0= u$, $r_1 = v$, $r_2 = w$ ($0 <u<1$, $0 <v<1$, $0 <w<1$, $u+v+w =1$). Since Alice's strategy is deterministic one, the three probability distributions $P_{0} (a|A_x)$, $P_{1} (a|A_x)$ and $P_{2} (a|A_x)$ must be equal to any three among $P_D^{00}$, $P_D^{01}$, $P_D^{10}$ and $P_D^{11}$. But any such combination will not satisfy the marginal probabilities $P(a|A_x)$ for Alice. So it is impossible to generate a LHV-LHS decomposition of the bi-unsteerable noisy Mermin family with hidden variables having dimension $3$ and with different deterministic distributions at Alice's side.

Let us try to generate a LHV-LHS decomposition of the bi-unsteerable noisy Mermin family  from Alice to Bob-Charlie in one sided device independent scenario with hidden variables having dimension $3$ and with different \textit{non-deterministic} distributions at Alice's side. We note that the noisy Mermin family is fully local for $V \leq \frac{1}{2}$ and it is two-way local, but not fully local for $ \frac{1}{2} < V \leq 1$. Hence from any decomposition of the noisy Mermin family in terms of fully
deterministic boxes or two-way local vertices, one may construct a LHV-LHS model of the bi-unsteerable noisy Mermin family as in Eq.(\ref{lhvlhs}) with different deterministic distributions at Alice's side, which does not require hidden variables of dimension more than 4 since there are only 4 possible different deterministic distributions given by Eq.(\ref{dba}) at Alice's side. Hence, a LHV-LHS model of the bi-unsteerable noisy Mermin family from Alice to Bob-Charlie in one sided device independent scenario with hidden variables of dimension $3$  can also be achieved by constructing a LHV-LHS model of the bi-unsteerable noisy Mermin family from Alice to Bob-Charlie in one sided device independent scenario with hidden variables of dimension $4$ with different deterministic distributions at Alice's side followed by taking equal joint probability distributions (having quantum realisations) at Bob-Charlie's side as common and making the corresponding distributions at Alice's side non-deterministic.

If the hidden variable dimension in the LHV-LHS decomposition of bi-unsteerable noisy Mermin family from Alice to Bob-Charlie in one sided device independent scenario can be reduced from $4$ to $3$, then noisy Mermin family can be decomposed in the following way:
\begin{equation}
\label{new1}
P_{MF}^{V} (abc|A_x B_ y C_z) = \sum_{\lambda=0}^{3} r_{\lambda} P_{\lambda} (a|A_x) P(b c |B_y, C_z,  \rho^{\lambda}_{BC}),
\end{equation}
where $P_{\lambda} (a|A_x)$ are different deterministic distributions
and any two of the four joint probability distributions $P(b c |B_y, C_z,  \rho^{\lambda}_{BC})$ 
are equal to each other; $0 < r_{\lambda} < 1$ for $\lambda$ = $0,1,2,3$; 
$\sum_{\lambda=0}^{3} r_{\lambda} = 1$. Then taking equal joint probability distributions 
$P(b c |B_y, C_z,  \rho^{\lambda}_{BC})$ at Bob-Charlie's side as common and making corresponding distribution at Alice's side non-deterministic will reduce the dimension of the hidden variable from $4$ to $3$. For example, let us consider
\begin{equation}
P(b c |B_y, C_z,  \rho^{0}_{BC}) = P(b c |B_y, C_z,  \rho^{2}_{BC}).
\end{equation}
Now in order to satisfy Alice's marginal given by Eq. (\ref{mar}), one must take $r_0$ = $r_1$ = $r_2$ = $r_3$ = $\frac{1}{4}$. 
Hence, the decomposition (\ref{new1}) can be written as, 
\begin{align}
\label{new2}
P_{MF}^{V} (abc|A_x B_ y C_z) =& q_0 \mathbb{P}_{0}(a|A_x) P(b c |B_y, C_z,  \rho^{0}_{BC}) 
 + \frac{1}{4} P_{1} (a|A_x) P(b c |B_y, C_z,  \rho^{1}_{BC}) 
 + \frac{1}{4} P_{3} (a|A_x) P(b c |B_y, C_z,  \rho^{3}_{BC}),
\end{align} 
where,
\begin{equation}
\mathbb{P}_{0}(a|A_x) = \frac{P_{0} (a|A_x)+  P_{2} (a|A_x)}{2},
\end{equation}
which is a non-deterministic distribution at Alice's side, and
\begin{equation}
q_0 = \frac{1}{2}.
\end{equation} 
The decomposition (\ref{new2}) represents a LHV-LHS model of bi-unsteerable noisy Mermin family  from Alice to Bob-Charlie in one sided device independent scenario
having different deterministic/non-deterministic distributions at Alice's side with the dimension of the hidden variable being $3$. Now in this protocol, if all the tripartite distributions $P_{MF}^{V} (abc|A_x B_ y C_z)$ are reproduced, quantum realizations of all the joint probability distributions $P(b c |B_y, C_z,  \rho^{\lambda}_{BC})$ are not possible for $ V > \frac{1}{\sqrt{5}}$ (for detailed calculations, see the Appendix \ref{a2}).

There are the following other cases in which the dimension of the hidden variable in the LHV-LHS decomposition of the bi-unsteerable noisy Mermin family from Alice to Bob-Charlie in one sided device independent scenario can be reduced from $4$ to $3$:
\begin{align}
& i) \hspace{0.3cm} P(b c |B_y, C_z,  \rho^{0}_{BC}) = P(b c |B_y, C_z,  \rho^{1}_{BC}); \nonumber \\ 
& ii) \hspace{0.3cm} P(b c |B_y, C_z,  \rho^{0}_{BC}) = P(b c |B_y, C_z,  \rho^{3}_{BC});  \nonumber  \\
& iii) \hspace{0.3cm} P(b c |B_y, C_z,  \rho^{1}_{BC}) = P(b c |B_y, C_z,  \rho^{2}_{BC});  \nonumber \\
& iv) \hspace{0.3cm} P(b c |B_y, C_z,  \rho^{1}_{BC}) = P(b c |B_y, C_z,  \rho^{3}_{BC});  \nonumber \\
& v) \hspace{0.3cm} P(b c |B_y, C_z,  \rho^{2}_{BC}) = P(b c |B_y, C_z,  \rho^{3}_{BC});  \nonumber 
\end{align} 
Now in cases $i)$ and $v)$, it can be shown that all the tripartite distributions $P_{MF}^{V} (abc|A_x B_ y C_z)$ is reproduced \textit{iff} $V=0$. On the other hand, in cases $ii)$, $iii)$ and $iv)$, following similar procedure adopted in Appendix \ref{a2} it can be shown that if all the tripartite distributions $P_{MF}^{V} (abc|A_x B_ y C_z)$ are reproduced, quantum realizations of all the joint probability distributions $P(b c |B_y, C_z,  \rho^{\lambda}_{BC})$ are not possible for $ V > \frac{1}{\sqrt{5}}$.

Hence, one can conclude that the LHV-LHS decomposition of bi-unsteerable noisy Mermin box  from Alice to Bob-Charlie in one sided device independent scenario cannot be realized with hidden variables having dimension $3$ for $V > \frac{1}{\sqrt{5}}$ with deterministic/non-deterministic distributions at Alice's side.\\

\textbf{Theorem 2.} \textit{The LHV-LHS decomposition of bi-unsteerable
noisy Mermin box from Alice to Bob-Charlie in one sided device independent scenario} cannot be realized with hidden variables having dimension $2$ or $1$ for $V > 0$.\\

\textit{Proof.} Now, let us try to generate a LHV-LHS decomposition of the bi-unsteerable noisy Mermin family  from Alice to Bob-Charlie in one sided device independent scenario with hidden variables of dimension $2$ having different deterministic distributions at Alice's side. In this case the noisy Mermin family can be decomposed in the following way:
\begin{equation}
P_{MF}^{V} (abc|A_x B_ y C_z) = \sum_{\lambda=0}^{1} r_{\lambda} P_{\lambda} (a|A_x) P(b c |B_y, C_z,  \rho^{\lambda}_{BC}).
\end{equation}
Here, $r_0=u$, $r_1=v$ ($0 <u<1$, $0 <v<1$, $u+v =1$). Since Alice's strategies are deterministic, the two probability distributions $P_{0} (a|A_x)$ and $P_{1} (a|A_x)$ must be equal to any two among $P_D^{00}$, $P_D^{01}$, $P_D^{10}$ and $P_D^{11}$. In order to satisfy the marginal probabilities for Alice, the only two possible choices of $P_{0} (a|A_x)$ and $P_{1} (a|A_x)$ are:\\
1) $P_D^{00}$ and $P_D^{01}$ with $u=v=\frac{1}{2}$\\
2) $P_D^{10}$ and $P_D^{11}$ with $u=v=\frac{1}{2}$.

Now, it can be easily checked that none of these two possible choices will satisfy all the tripartite joint probability distributions $P_{MF}^{V} (abc|A_x B_ y C_z)$ for $V>0$ (for detailed calculations, see the Appendix \ref{a3}). It is, therefore, impossible to generate a LHV-LHS decomposition of the bi-unsteerable noisy Mermin family from Alice to Bob-Charlie in one sided device independent scenario with hidden variables of dimension $2$ having different deterministic distributions at Alice's side. 

Now, let us try to generate a LHV-LHS decomposition of the bi-unsteerable noisy Mermin family  from Alice to Bob-Charlie in one sided device independent scenario with hidden variables having dimension $2$ and with different \textit{non-deterministic} distributions at Alice's side. As noted earlier, this  can also be achieved by constructing a LHV-LHS model of the bi-unsteerable noisy Mermin family  from Alice to Bob-Charlie in one sided device independent scenario with hidden variables of dimension $4$ or $3$ having different deterministic distributions at Alice's side followed by taking equal joint probability distributions (having quantum realizations) at Bob-Charlie's side as common and making the corresponding distributions at Alice's side non-deterministic.

It has already been shown that it is impossible to generate a LHV-LHS decomposition of the bi-unsteerable noisy Mermin family  from Alice to Bob-Charlie in one sided device independent scenario with hidden variables having dimension $3$ and with different deterministic distributions at Alice's side. Hence, there is no scope to reduce the hidden variable dimension from $3$ to $2$ in the LHV-LHS decomposition of bi-unsteerable noisy Mermin family.

Now, if the hidden variable dimension in the LHV-LHS decomposition of bi-unsteerable noisy Mermin family  from Alice to Bob-Charlie in one sided device independent scenario can be reduced from $4$ to $2$, then noisy Mermin family can be decomposed in the following way:
\begin{equation}
\label{new11}
P_{MF}^{V} (abc|A_x B_ y C_z) = \sum_{\lambda=0}^{3} r_{\lambda} P_{\lambda} (a|A_x) P(b c |B_y, C_z,  \rho^{\lambda}_{BC}),
\end{equation}
where $P_{\lambda} (a|A_x)$ are different deterministic distributions;
and either any three of the four joint probability distributions $P(b c |B_y, C_z,  \rho^{\lambda}_{BC})$ 
are equal to each other or there exists two sets each containing two equal joint probability distributions $P(b c |B_y, C_z,  \rho^{\lambda}_{BC})$; $0 < r_{\lambda} < 1$ for $\lambda$ = $0,1,2,3$; 
$\sum_{\lambda=0}^{3} r_{\lambda} = 1$. Then, as described earlier, taking equal joint probability distributions 
$P(b c |B_y, C_z,  \rho^{\lambda}_{BC})$ at Bob-Charlie's side as common and making corresponding distribution at Alice's side non-deterministic will reduce the dimension of the hidden variable from $4$ to $2$.

There are the following seven cases in which the dimension of the hidden variable in the LHV-LHS decomposition of bi-unsteerable noisy Mermin family  from Alice to Bob-Charlie in one sided device independent scenario can be reduced from $4$ to $2$: 
\begin{center}
$ P(b c |B_y, C_z,  \rho^{0}_{BC}) = P(b c |B_y, C_z,  \rho^{1}_{BC}) = P(b c |B_y, C_z,  \rho^{2}_{BC}); $ \\ 
$ P(b c |B_y, C_z,  \rho^{0}_{BC}) = P(b c |B_y, C_z,  \rho^{1}_{BC}) = P(b c |B_y, C_z,  \rho^{3}_{BC}); $ \\ 
$ P(b c |B_y, C_z,  \rho^{0}_{BC}) = P(b c |B_y, C_z,  \rho^{2}_{BC}) = P(b c |B_y, C_z,  \rho^{3}_{BC}); $ \\ 
$ P(b c |B_y, C_z,  \rho^{1}_{BC}) = P(b c |B_y, C_z,  \rho^{2}_{BC}) = P(b c |B_y, C_z,  \rho^{3}_{BC}); $ \\
$P(b c |B_y, C_z,  \rho^{0}_{BC})$ = $P(b c |B_y, C_z,  \rho^{1}_{BC})$ as well as $P(b c |B_y, C_z,  \rho^{2}_{BC})$ = $P(b c |B_y, C_z,  \rho^{3}_{BC})$;\\
$P(b c |B_y, C_z,  \rho^{0}_{BC})$ = $P(b c |B_y, C_z,  \rho^{2}_{BC})$ as well as $P(b c |B_y, C_z,  \rho^{1}_{BC})$ = $P(b c |B_y, C_z,  \rho^{3}_{BC})$; \\
$P(b c |B_y, C_z,  \rho^{0}_{BC})$ = $P(b c |B_y, C_z,  \rho^{3}_{BC})$ as well as $P(b c |B_y, C_z,  \rho^{1}_{BC})$ = $P(b c |B_y, C_z,  \rho^{2}_{BC})$;
\end{center}
Now in any of these possible cases, considering arbitrary joint probability distributions 
$P(b c |B_y, C_z,  \rho^{\lambda}_{BC})$ at Bob-Charlie's side (without considering any constraint), it can be shown that all the tripartite distributions $P_{MF}^{V} (abc|A_x B_ y C_z)$ are not reproduced simultaneously for $V>0$. 
Hence, this also holds when the boxes $P_{\lambda}^{Sv} (b c|y z)$ satisfy 
NS principle as well as have quantum realizations.

It can be checked that the noisy Mermin box is non-product across all three bipartite cuts for 
any $V>0$. It is, therefore, impossible to generate a LHV-LHS decomposition of the bi-unsteerable
noisy Mermin box ($0 < V \leq \frac{1}{\sqrt{2}}$) from Alice to Bob-Charlie in one sided device independent scenario with hidden variables having dimension $1$.

Hence, one can conclude that the LHV-LHS decomposition of bi-unsteerable
noisy Mermin box from Alice to Bob-Charlie in one sided device independent scenario cannot be realized with hidden variables having dimension $2$ or $1$ for $V>0$.\\

Theorem 2 implies the following.\\

\textbf{Corollary 1.} \textit{The bi-unsteerable noisy Mermin family demonstrates
super-bi-unsteerablity  from Alice to Bob-Charlie in one sided device independent scenario  for $0 < V \leq \frac{1}{\sqrt{2}}$.}\\

\textit{Proof.} The bi-unsteerable noisy Mermin family ($0 < V \leq \frac{1}{\sqrt{2}}$) can be reproduced by appropriate measurements on the quantum state in $\mathbb{C}^{2}\otimes\mathbb{C}^{2}\otimes\mathbb{C}^{2}$ (given by Eq.(\ref{GHZ})). On the other hand, we have shown that the bi-unsteerable noisy Mermin family ($0 < V \leq \frac{1}{\sqrt{2}}$) can be simulated with LHV at Alice's side and LHS at Bob-Charlie's side with the minimum dimension of the hidden variable being greater than $2$. The bi-unsteerable noisy Mermin family ($0 < V \leq \frac{1}{\sqrt{2}}$), therefore, demonstrates super-bi-unsteerability  from Alice to Bob-Charlie in one sided device independent scenario.\\

The above Corollary implies the following theorem.\\

\textbf{Theorem 3.} \textit{The bi-unsteerable noisy Mermin family demonstrates genuine
super-bi-unsteerablity in one sided device independent scenario  for $0 < V \leq \frac{1}{\sqrt{2}}$.}\\

\textit{Proof.} Since noisy Mermin family is invariant under permutations of parties, the bi-unsteerable noisy Mermin family demonstrates
super-bi-unsteerablity from Bob to Alice-Charlie and from Charlie to Alice-Bob in one sided device independent scenario for $0 < V \leq \frac{1}{\sqrt{2}}$. Hence, the bi-unsteerable noisy Mermin family demonstrates genuine
super-bi-unsteerablity in one sided device independent scenario  for $0 < V \leq \frac{1}{\sqrt{2}}$.\\

Now, we consider quantum correlations that belong to the noisy Svetlichny family defined as
\begin{equation}
P_{SvF}^{V} (abc|A_x B_ y C_z) = \frac{2 + (-1)^{a \oplus b \oplus c \oplus  xy \oplus yz \oplus xz} \sqrt{2} V}{16}  
\end{equation}
where $0 < V \leq 1$. Since the noisy Svetlichny family has nonzero Svetlichny strength for any $V > 0$, the quantum simulation of these correlations by using a $2 \otimes 2 \otimes 2$ quantum state necessarily requires genuine quantum discord  \cite{mp1, mp2} in the state \cite{mp4, mp10}. Following the similar argument presented earlier in case noisy Mermin family, it can be stated that the bi-unsteerable noisy Svetlichny family demonstrates genuine super-bi-unsteerablity in one sided device independent scenario for $0 < V \leq \frac{1}{\sqrt{2}}$ \cite{sttr2}.

\section{Genuine Quantumess of tripartite correlations as captured by ``genuine super-bi-unsteerablity"}

Note that the dimension of the hidden variable needed to simulate the LHV-LHS model of the bi-unsteerable noisy Mermin family across any possible bipartition in one sided device independent scenario in the range $V > \frac{1}{\sqrt{5}}$ must be greater than $3$. On the other hand, that in the range $V > 0$ must be greater than $2$. Hence, the  genuinely super-bi-unsteerable noisy Mermin family certifies genuine quantumness of the $2 \otimes 2 \otimes 2$ dimensional resource reproducing it in the range $0 < V \leq \frac{1}{\sqrt{2}}$. For example, the  genuinely super-bi-unsteerable noisy Mermin family in the range $0 < V \leq \frac{1}{\sqrt{2}}$ characterizes the genuine quantumness of the state given by Eq.(\ref{GHZ}). The genuinely super-bi-unsteerable noisy Mermin family in the range $\frac{1}{\sqrt{5}} < V \leq \frac{1}{\sqrt{2}}$ also certifies genuine quantumness of the $3 \otimes 2 \otimes 2$ dimensional resource reproducing it. For example, consider that the three spatially separated parties (say, Alice, Bob and Charlie) share the following $3 \otimes 2 \otimes 2$ quantum state:
\begin{equation}
\label{qutrt}
\rho_2 = V | GHZ \rangle \langle GHZ | + (1-V) |2\rangle \langle 2| \otimes \frac{\mathbb{I}_2}{2} \otimes \frac{\mathbb{I}_2}{2}
\end{equation}
where $| GHZ \rangle = \frac{1}{\sqrt{2}} (|000 \rangle + |111\rangle)$; $0 <V \leq 1$; $|0\rangle$, $|1\rangle$ and $|2\rangle$ form an orthonormal basis in the Hilbert space in $\mathcal{C}^3$;  $|0\rangle$ and $|1\rangle$ form an orthonormal basis in the Hilbert space in $\mathcal{C}^2$; $\mathbb{I}_2 = |0\rangle \langle 0| + |1\rangle \langle 1|$. If Alice, Bob and Charlie perform appropriate measurements on the state given in Eq.(\ref{qutrt}), the noisy Mermin family can be reproduced (for detailed calculations, see the Appendix \ref{a4}). Hence, the genuinely super-bi-unsteerable noisy Mermin family in the range $\frac{1}{\sqrt{5}} < V \leq \frac{1}{\sqrt{2}}$ characterizes the genuine quantumness the $3 \otimes 2 \otimes 2$ state given by Eq.(\ref{qutrt}).

The notion of genuine tripartite quantum discord has been defined in a tripartite quantum state to capture the genuine quantumness of separable 
states  \cite{mp1}. Genuine tripartite quantum discord becomes zero \emph{iff} there exists a bipartite cut of the tripartite system such that no quantum correlation exist between the two parts \cite{mp2}. It is well-known that a bipartite quantum state has no (Alice to Bob) quantum discord \emph{iff} it can be written in the classical-quantum (CQ) state form, $\rho_{CQ}=\sum_ip_i|i\rangle^{A}\langle i|\otimes\rho_i^B$ \cite{cq}.

The tripartite classical-quantum state is defined  as follows.

\textbf{Definition 4:} \textit{A fully separable tripartite state has a classical-quantum state form with respect to the bipartite cut $A$ versus $BC$ if it can be decomposed as 
\begin{equation}
\rho^{A|BC}_{CQ}=\sum_ip_i |i\rangle^{A}\langle i| \otimes \rho^{B}_i \otimes \rho^{C}_i, \label{cqBC}
\end{equation}
where $\{|i\rangle^{A}\langle i|\}$ is some orthonormal basis of Alice's Hilbert space $\mathcal{H}_A$.}

The tripartite quantum states which have the classical-quantum state form given above do not have nonzero genuine quantum discord since subsystem $A$ is always classically correlated with $B$ and $C$ subsystems. Now, Consider tripartite boxes arising from three-qubit classical-quantum states which have the form as given in Eq.(\ref{cqBC}) with $i = 0, 1$. The correlations obtained from this state can  manifestly  be simulated  by presharing classical random variable $\lambda$ of dimension  $2$. Hence, the states given by Eq.(\ref{cqBC}) represent a  family of states that do not demonstrate super-bi-unsteerability from $A$ to $BC$. This implies that for any three-qubit state which do not have genuine quantumness, there exists a bipartite cut in which it is not super-bi-unsteerable. One can, therefore, conclude that genuine nonclassicality of bi-unsteerable correlations (produced from three-qubit states) \cite{mp4} is necessary for implying genuine super-bi-unsteerability.

\section{Discussion and Conclusions}

In the present work we have introduced the notion of super-bi-unsteerability by showing that there are certain bi-unsteerable correlations whose simulation with LHV-LHS model requires preshared randomness with dimension higher than the local Hilbert space dimension of the quantum systems (reproducing the given bi-unsteerable correlations) at the untrusted party's side. The super-bi-unsteerability of the noisy Mermin family has been demonstrated in the present study.

In   Ref.   \cite{sl5},  the   authors  have  shown   that  the nonclassicality of a  family of bipartite local correlations  in the Bell-CHSH scenario can be characterized by  superlocality. Extending this  approach, it has been shown that the nonclassicality  in the related steering scenario can also be pointed out by  the notion of super-unsteerability \cite{sus} of certain bipartite unsteerable correlations. The notion of superlocality of bipartite local correlations has also been generalized to demonstrate superlocality of multipartite boxes \cite{mp10}.  Motivated by this, in the present paper, we generalize the concept of super-unsteerability in the tripartite scenario and define the notion of ``super-bi-unsteerability" and ``genuine super-bi-unsteerability" in the context of tripartite bi-unsteerable correlations.

Before concluding, we note that nonlocality or steerability of any  correlation in QM or in any convex operational theory can be characterized by the non-zero communication cost that must be supplemented with preshared randomness in order to simulate the correlations. The question of an analogous operational characterization of genuine quantumness of bi-unsteerable tripartite correlations has been addressed here, and associated with genuine super-bi-unsteerability. %Since, every tripartite superlocal correlations \cite{mp10} are super-bi-unsteerable, but the converse not being true, finding out tripartite correlations which are super-bi-unsteerable, but not superlocal is an interesting area for future studies.

 In the present study we have restricted ourselves to one sided device independent scenario. Investigating genuine super-bi-unsteerability in two sided device independent scenario is an interesting area for future studies. It is worth to be studied whether there exists any quantum information theoretic application of genuine super-bi-unsteerability.

\section{ACKNOWLEDGEMENTS}
DD acknowledges the financial support from University Grants Commission (UGC), Government of India. BB acknowledges the financial support from Department of Science and Technology (DST), Government of India. CJ is  thankful to Prof. R. Srikanth for fruitful discussions.  AM acknowledges support from the CSIR project 09/093(0148)/2012-EMR-I.

\appendix

\section{Demonstrating super-unsteerability of each joint probability distribution at Bob-Charlie's side $P(b c |B_y, C_z,  \rho^{\lambda}_{BC})$ produced from the LHS of the LHV-LHS decomposition of noisy Mermin family when $0 < V \leq \frac{1}{2}$}	\label{a1}

The correlation belonging to noisy Mermin family can be written as
\begin{equation}
P_{MF}^{V} (abc|A_x B_ y C_z) = \sum_{\lambda=0}^{3} r_{\lambda} P_{\lambda} (a|A_x) P(b c |B_y, C_z,  \rho^{\lambda}_{BC}),
\end{equation}
where $r_0$ = $r_1$ = $r_2$ = $r_3$ = $\frac{1}{4}$, and \\
$P_{0} (a|A_x)$ = $P_D^{00}$, $P_{1} (a|A_x)$ = $P_D^{01}$, $P_{2} (a|A_x)$ = $P_D^{10}$, $P_{3} (a|A_x)$ = $P_D^{11}$.

Now,
 \begin{equation}
 P(b c |B_y, C_z,  \rho^{0}_{BC}) = \bordermatrix{
\frac{bc}{yz} & 00 & 01 & 10 & 11 \cr
00 & \frac{1+V}{4} & \frac{1-V}{4} & \frac{1-V}{4} & \frac{1+V}{4} \cr
01 & \frac{1+V}{4} & \frac{1-V}{4} & \frac{1-V}{4} & \frac{1+V}{4} \cr
10 & \frac{1+V}{4} & \frac{1-V}{4} & \frac{1-V}{4} & \frac{1+V}{4} \cr
11 & \frac{1-V}{4} & \frac{1+V}{4} & \frac{1+V}{4} & \frac{1-V}{4} } ,
\end{equation}
where each row and column corresponds to a fixed measurement $(yz)$ and a fixed outcome $(bc)$ respectively. This correlation can be written as
\begin{equation}
P(b c |B_y, C_z,  \rho^{0}_{BC}) = \sum_{\lambda=0}^{3} q^{0}_{\lambda} P^{0}_{\lambda} (b|B_y) P^{0}(c | C_z,  \rho^{\lambda}_{C}),
\end{equation}
where $q^{0}_0$ = $q^{0}_1$ = $q^{0}_2$ = $q^{0}_3$ = $\frac{1}{4}$, and \\
$P^{0}_{0} (b|B_y)$ = $P_D^{00}$, $P^{0}_{1}(b|B_y)$ = $P_D^{01}$, $P^{0}_{2} (b|B_y)$ = $P_D^{10}$, $P^{0}_{3}(b|B_y)$ = $P_D^{11}$.

Now,
 \begin{equation}
P^{0}(c | C_z,  \rho^{0}_{C}) = \bordermatrix{
\frac{c}{z} & 0 & 1 \cr
0 & \frac{1+2V}{2} & \frac{1-2V}{2} \cr
1 & \frac{1}{2} & \frac{1}{2} },
\end{equation}
where each row and column corresponds to a fixed measurement $(z)$ and a fixed outcome $(c)$ respectively. Now,  $0 \leq P^{0}(c | C_z,  \rho^{0}_{C}) \leq 1$ $\forall c,z$, which implies that $0 < V \leq \frac{1}{2}$.

This probability distribution at Charlie's side can be reproduced by performing the projective qubit measurements of the observables corresponding to the operators: $C_0 = \sigma_y$, $C_1 = - \sigma_x$ on the state given by
\begin{equation}
\label{state11}
|\psi^0_{C}\rangle = \cos \theta | 0 \rangle + e^{i \phi_0} \sin \theta |1 \rangle,
\end{equation}
where, $\phi_0 = \dfrac{\pi}{2}$; $\sin 2 \theta = 2 V$; $|0\rangle$ and $|1\rangle$ are the eigenstates of $\sigma_z$ corresponding to the eigenvalues $+1$ and $-1$ respectively.

 \begin{equation}
P^{0}(c | C_z,  \rho^{1}_{C}) = \begin{pmatrix}
 \frac{1-2V}{2} & \frac{1+2V}{2} \\
 \frac{1}{2} & \frac{1}{2}
 \end{pmatrix}.
\end{equation}
Now,  $0 \leq P^{0}(c | C_z,  \rho^{1}_{C}) \leq 1$ $\forall c,z$, which implies that $0 < V \leq \frac{1}{2}$.

This probability distribution at Charlie's side can be reproduced by performing the projective qubit measurements of the observables corresponding to the operators: $C_0 = \sigma_y$, $C_1 = - \sigma_x$ on the state given by
\begin{equation}
\label{state12}
|\psi^1_{C}\rangle = \cos \theta | 0 \rangle - e^{i \phi_1} \sin \theta |1 \rangle,
\end{equation}
where, $\phi_1 = \dfrac{\pi}{2}$; $\sin 2 \theta = 2 V$.

 \begin{equation}
P^{0}(c | C_z,  \rho^{2}_{C}) = \begin{pmatrix}
 \frac{1}{2} & \frac{1}{2} \\
 \frac{1+2V}{2} & \frac{1-2V}{2}
 \end{pmatrix}.
\end{equation}
Now,  $0 \leq P^{0}(c | C_z,  \rho^{2}_{C}) \leq 1$ $\forall c,z$, which implies that $0 < V \leq \frac{1}{2}$.

This probability distribution at Charlie's side can be reproduced by performing the projective qubit measurements of the observables corresponding to the operators: $C_0 = \sigma_y$, $C_1 = - \sigma_x$ on the state given by
\begin{equation}
\label{state12}
|\psi^2_{C}\rangle = \cos \theta | 0 \rangle - e^{i \phi_2} \sin \theta |1 \rangle,
\end{equation}
where, $\phi_2 = 0$; $\sin 2 \theta = 2 V$.

\begin{equation}
P^{0}(c | C_z,  \rho^{3}_{C}) = \begin{pmatrix}
 \frac{1}{2} & \frac{1}{2} \\
 \frac{1-2V}{2} & \frac{1+2V}{2}
 \end{pmatrix}.
\end{equation}
Now,  $0 \leq P^{0}(c | C_z,  \rho^{3}_{C}) \leq 1$ $\forall c,z$, which implies that $0 < V \leq \frac{1}{2}$.

This probability distribution at Charlie's side can be reproduced by performing the projective qubit measurements of the observables corresponding to the operators: $C_0 = \sigma_y$, $C_1 = - \sigma_x$ on the state given by
\begin{equation}
\label{state12}
|\psi^3_{C}\rangle = \cos \theta | 0 \rangle + e^{i \phi_3} \sin \theta |1 \rangle,
\end{equation}
where, $\phi_3 = 0$; $\sin 2 \theta = 2 V$.

Hence, one can state that $P(b c |B_y, C_z,  \rho^{0}_{BC})$ can be expressed with a LHV-LHS decomposition having hidden variables of  dimension $4$ with different deterministic distributions at Bob's side in the range  $0 < V \leq \frac{1}{2}$.

Now, let us try to generate a LHV-LHS decomposition of $P(b c |B_y, C_z,  \rho^{0}_{BC})$ with hidden variables having dimension $2$ and  with different deterministic distributions at Bob's side. Before proceeding, we want to mention that in case of $P(b c |B_y, C_z,  \rho^{0}_{BC})$, all the marginal probability distributions of Bob and Charlie are maximally mixed:
\begin{equation}
\label{mbc}
P(b|B_y, \rho^{0}_{BC}) = P(c|C_z, \rho^{0}_{BC}) = \frac{1}{2} \forall b,c,y,z
\end{equation} 
Now, in this case the unsteerable box $P(b c |B_y, C_z,  \rho^{0}_{BC})$ can be decomposed in the following way:
\begin{equation}
P(b c |B_y, C_z,  \rho^{0}_{BC})= \sum_{\lambda=0}^{1} q^{0}_{\lambda} P^{0}_{\lambda} (b|B_y) P^{0}(c |C_z,  \rho^{\lambda}_{C}).
\end{equation}
Here, $q^{0}_0=e$, $q^{0}_1=f$ ($0 <e<1$, $0 <f<1$, $e+f =1$). Since Bob's strategy is deterministic one, the two probability distributions $P^{0}_{0} (b|B_y)$ and $P^{0}_{1}(b|B_y)$ must be equal to any two among $P_D^{00}$, $P_D^{01}$, $P_D^{10}$ and $P_D^{11}$. In order to satisfy the marginal probabilities for Bob $P(b|B_y, \rho^{0}_{BC})$, the only two possible choices of $P^{0}_{0} (b|B_y)$ and $P^{0}_{1} (b|B_y)$ are:\\
1) $P_D^{00}$ and $P_D^{01}$ with $e=f=\frac{1}{2}$\\
2) $P_D^{10}$ and $P_D^{11}$ with $e=f=\frac{1}{2}$.\\
Now, it can be easily checked that none of these two possible choices will satisfy all the joint probability distributions $P(b c |B_y, C_z,  \rho^{0}_{BC})$ simultaneously. It is, therefore, impossible to generate a LHV-LHS decomposition of $P(b c |B_y, C_z,  \rho^{0}_{BC})$ with hidden variables having dimension $2$  and  with different deterministic distributions at Bob's side.

Now, we will show that it is impossible to generate a LHV-LHS decomposition of $P(b c |B_y, C_z,  \rho^{0}_{BC})$ with hidden variables having dimension $2$  and with \textit{deterministic or non-deterministic} distributions at Bob's side. Before proceeding we note that from any
decomposition of the unsteerable (as well as local) box $P(b c |B_y, C_z,  \rho^{0}_{BC})$ ($0 < V \leq \frac{1}{2}$) in terms of deterministic boxes (\ref{DB}), one may derive a LHV-LHS model with different deterministic distributions at Bob's side, which does not require Bob to preshare the hidden variable
of dimension more than $4$ \cite{sl1} since there are only $4$ possible different deterministic distributions given by Eq. (\ref{db1}) at Bob's side. Hence, a LHV-LHS model with hidden  variable of dimension $2$ of the unsteerable box $P(b c |B_y, C_z,  \rho^{0}_{BC})$ ($0 < V \leq \frac{1}{2}$) can be achieved by constructing a LHV-LHS model of the unsteerable box $P(b c |B_y, C_z,  \rho^{0}_{BC})$ ($0 < V \leq \frac{1}{2}$) with hidden variable of dimension $3$ or $4$ with different deterministic distributions at Bob's side followed by taking equal probability distributions at Charlie's side as common and making the 
corresponding distributions at Bob's side non-deterministic.

Let us try to produce a LHV-LHS decomposition of $P(b c |B_y, C_z,  \rho^{0}_{BC})$ with hidden variables having dimension $3$ and  with different deterministic distributions at Bob's side.
In this case the unsteerable box $P(b c |B_y, C_z,  \rho^{0}_{BC})$ can be decomposed in the following way:
\begin{equation}
P(b c |B_y, C_z,  \rho^{0}_{BC})= \sum_{\lambda=0}^{2} q^{0}_{\lambda} P^{0}_{\lambda} (b|B_y) P^{0}(c |C_z,  \rho^{\lambda}_{C}).
\end{equation}
Here, $q^{0}_0= e$, $q^{0}_1 = f$, $q^{0}_2 = g$ ($0 <e<1$, $0 <f<1$, $0 <g<1$, $e+f+g =1$). Since Bob's strategy is deterministic one, the three probability distributions $P^{0}_{0} (b|B_y)$, $P^{0}_{1} (b|B_y)$ and $P_{2} (b|B_y)$ must be equal to any three among $P_D^{00}$, $P_D^{01}$, $P_D^{10}$ and $P_D^{11}$. But any such combination will not satisfy the marginal probabilities $P(b|B_y, \rho^{0}_{BC})$ for Bob. So it is impossible to generate a LHV-LHS decomposition of $P(b c |B_y, C_z,  \rho^{0}_{BC})$ with hidden variables having dimension $3$ and  with different deterministic distributions at Bob's side.

Therefore, in order to simulate the LHV-LHS decomposition of $P(b c |B_y, C_z,  \rho^{0}_{BC})$ with different deterministic distributions at Bob's side, Bob has to share the hidden variables of dimension $4$.  

Suppose the unsteerable box $P(b c |B_y, C_z,  \rho^{0}_{BC})$ can be decomposed in the following way:
\begin{equation}
\label{neww1}
P(b c |B_y, C_z,  \rho^{0}_{BC}) = \sum_{\lambda=0}^{3} q^{0}_{\lambda} P^{0}_{\lambda} (b|B_y) P^{0}(c |C_z,  \rho^{\lambda}_{C}),
\end{equation} 
where $P^{0}_{\lambda} (b|B_y)$ are different deterministic distributions
and either any three of the four probability distributions $P^{0}(c |C_z,  \rho^{\lambda}_{C})$ 
are equal to each other, or there exists two sets each containing two equal probability 
distributions $P^{0}(c |C_z,  \rho^{\lambda}_{C})$; $0 < q^{0}_{\lambda} < 1$ for $\lambda$ = $0,1,2,3$; 
$\sum_{\lambda=0}^{3} q^{0}_{\lambda} = 1$. Then taking equal probability distributions 
$P^{0}(c |C_z,  \rho^{\lambda}_{C})$ at Charlie's side as common and making corresponding distribution 
at Bob's side non-deterministic will reduce the dimension of the hidden variable from $4$ to $2$.

Now in order to satisfy Bob's marginal given by Eq.(\ref{mbc}), one must take $q^{0}_0$ = $q^{0}_1$ = $q^{0}_2$ = $q^{0}_3$ = $\frac{1}{4}$. It can be easily checked that for all possible cases, in which the hidden variable dimension in the LHV-LHS decomposition (\ref{neww1}) can be reduced from $4$ to $2$, all the joint probability distributions $P(b c |B_y, C_z,  \rho^{0}_{BC})$ are not satisfied simultaneously for $V>0$. This can be checked considering arbitrary probability distributions 
$P^{0}(c |C_z,  \rho^{\lambda}_{C})$ at Charlie's side (without considering any constraint). Hence, this also follows when the probability distributions 
$P^{0}(c |C_z,  \rho^{\lambda}_{C})$ at Charlie's side has quantum realisations. It is, therefore, impossible to reduce the dimension from $4$ to $2$ in the LHV-LHS decomposition (\ref{neww1}) of $P(b c |B_y, C_z,  \rho^{0}_{BC})$.

It can be checked that the joint probability distribution $P(b c |B_y, C_z,  \rho^{0}_{BC})$ is non-product. It is, therefore, impossible to generate a LHV-LHS decomposition of the joint probability distribution $P(b c |B_y, C_z,  \rho^{0}_{BC})$ with hidden variables having dimension $1$.

Hence, one can conclude that the LHV-LHS decomposition of
$P(b c |B_y, C_z,  \rho^{0}_{BC})$ cannot be realized with hidden variables having dimension $2$ or $1$.

Now, as stated before, the joint probability distribution $P(b c |B_y, C_z,  \rho^{0}_{BC})$ at Bob and Charlie's side can be reproduced by performing the projective qubit measurements of the observables corresponding to the operators $B_0 = \sigma_y$, $B_1 = - \sigma_x$; and $C_0 = \sigma_y$, $C_1 = - \sigma_x$ on the $2 \otimes 2$ quantum state given by
\begin{equation}
\label{state1}
| \psi_0 \rangle = \cos \theta |00 \rangle - \dfrac{1+i}{\sqrt{2}} \sin \theta |11 \rangle,
\end{equation}
$0 \leq \theta \leq \frac{\pi}{4}$ and $sin 2 \theta = \sqrt{2} V$; $|0\rangle$ and $|1\rangle$ are the eigenstates of $\sigma_z$ corresponding to the eigenvalues $+1$ and $-1$ respectively.

We have shown that the unsteerable box $P(b c |B_y, C_z,  \rho^{0}_{BC})$ ($0 < V \leq \frac{1}{2}$) can be simulated with LHV-LHS model with the minimum dimension of the hidden variable being greater than $2$. On the other hand, $P(b c |B_y, C_z,  \rho^{0}_{BC})$ can be simulated by appropriate measurement on $2 \otimes 2$ quantum system. Hence, one can state that the unsteerable box $P(b c |B_y, C_z,  \rho^{0}_{BC})$ demonstrates super-unsteerablity for $0 < V \leq \frac{1}{2}$.

In a similar way as described above, it can be shown that $P(b c |B_y, C_z,  \rho^{1}_{BC})$, $P(b c |B_y, C_z,  \rho^{2}_{BC})$ and $P(b c |B_y, C_z,  \rho^{3}_{BC})$ also demonstrate super-unsteerablity for $0 < V \leq \frac{1}{2}$.

\section{Reducing the dimension of the hidden variable from $4$ to $3$ in the LHV-LHS decomposition of the bi-unsteerable noisy Mermin family in the bipartition $A-BC$ in one sided device independent scenario} \label{a2}
Consider that the noisy Mermin family can be decomposed in the following way:
\begin{equation}
\label{aa1}
P_{MF}^{V} (abc|A_x B_ y C_z) = \sum_{\lambda=0}^{3} r_{\lambda} P_{\lambda} (a|A_x) P(b c |B_y, C_z,  \rho^{\lambda}_{BC}),
\end{equation}
where without any loss of generality let us assume that $P_{0} (a|A_x)= P_D^{00}$, $P_{1} (a|A_x)= P_D^{01}$, $P_{2} (a|A_x)= P_D^{10}$ and $P_{3} (a|A_x)= P_D^{11}$; and also assume that
$P(b c |B_y, C_z,  \rho^{0}_{BC}) = P(b c |B_y, C_z,  \rho^{2}_{BC})$.
Now in order to satisfy Alice's marginal given by Eq. (\ref{mar}), one must take $r_0$ = $r_1$ = $r_2$ = $r_3$ = $\frac{1}{4}$. 
Hence, the decomposition (\ref{aa1}) can be written as, 
\begin{align}
P_{MF}^{V} (abc|A_x B_ y C_z) =& q_0 \mathbb{P}_{0}(a|A_x) P(b c |B_y, C_z,  \rho^{0}_{BC}) 
 + \frac{1}{4} P_{1} (a|A_x) P(b c |B_y, C_z,  \rho^{1}_{BC}) 
 + \frac{1}{4} P_{3} (a|A_x) P(b c |B_y, C_z,  \rho^{3}_{BC}),
\label{aa}
\end{align} 
where,
\begin{equation}
\mathbb{P}_{0}(a|A_x) = \frac{P_{0} (a|A_x)+  P_{2} (a|A_x)}{2},
\end{equation}
which is a non-deterministic distribution at Alice's side, and
\begin{equation}
q_0 = \frac{1}{2}.
\end{equation} 
The decomposition (\ref{aa}) represents a LHV-LHS model of bi-unsteerable noisy Mermin family from Alice to Bob-Charlie in one sided device independent scenario
having different deterministic/non-deterministic distributions at Alice's side with the dimension of the hidden variable being $3$.

Now equating left hand side of Eq.(\ref{aa}) with its right hand side, we obtain the following unique  solution for the joint probability distributions at Bob-Charlie's side, 
\begin{equation}
 P(b c |B_y, C_z,  \rho^{0}_{BC}) = \bordermatrix{
\frac{bc}{yz} & 00 & 01 & 10 & 11 \cr
00 & \frac{1}{4} & \frac{1}{4} & \frac{1}{4} & \frac{1}{4} \cr
01 & \frac{1+V}{4} & \frac{1-V}{4} & \frac{1-V}{4} & \frac{1+V}{4} \cr
10 & \frac{1+V}{4} & \frac{1-V}{4} & \frac{1-V}{4} & \frac{1+V}{4} \cr
11 & \frac{1}{4} & \frac{1}{4} & \frac{1}{4} & \frac{1}{4} } ,
\label{n1}
\end{equation}
\begin{equation}
 P(b c |B_y, C_z,  \rho^{1}_{BC}) = \bordermatrix{
\frac{bc}{yz} & 00 & 01 & 10 & 11 \cr
00 & \frac{1-2V}{4} & \frac{1+2V}{4} & \frac{1+2V}{4} & \frac{1-2V}{4} \cr
01 & \frac{1-V}{4} & \frac{1+V}{4} & \frac{1+V}{4} & \frac{1-V}{4} \cr
10 & \frac{1-V}{4} & \frac{1+V}{4} & \frac{1+V}{4} & \frac{1-V}{4} \cr
11 & \frac{1+2V}{4} & \frac{1-2V}{4} & \frac{1-2V}{4} & \frac{1+2V}{4} } ,
\label{n2}
\end{equation}
and 
\begin{equation}
 P(b c |B_y, C_z,  \rho^{3}_{BC}) = \bordermatrix{
\frac{bc}{yz} & 00 & 01 & 10 & 11 \cr
00 &\frac{1+2V}{4} & \frac{1-2V}{4} & \frac{1-2V}{4} & \frac{1+2V}{4} \cr
01 & \frac{1-V}{4} & \frac{1+V}{4} & \frac{1+V}{4} & \frac{1-V}{4} \cr
10 & \frac{1-V}{4} & \frac{1+V}{4} & \frac{1+V}{4} & \frac{1-V}{4} \cr
11 & \frac{1-2V}{4} & \frac{1+2V}{4} & \frac{1+2V}{4} & \frac{1-2V}{4} } .
\label{n3}
\end{equation}
Now from the the necessary and sufficient condition for bipartite correlations to have quantum realisations \cite{nsqr}, it can be shown that the bipartite correlations (\ref{n1}), (\ref{n2}) and (\ref{n3}) will have quantum realizations \textit{iff} $V \leq \frac{1}{\sqrt{5}}$. Hence, the decomposition (\ref{aa1}) is not a LHV-LHS decomposition of noisy Mermin family for $V > \frac{1}{\sqrt{5}}$. Hence, in this case the dimension of the hidden variable in the LHV-LHS decomposition (\ref{aa1}) of the bi-unsteerable noisy Mermin family from Alice to Bob-Charlie in one sided device independent scenario cannot be reduced from $4$ to $3$ for $V > \frac{1}{\sqrt{5}}$.

\section{Demonstrating impossibility to have a LHV-LHS decomposition of the bi-unsteerable noisy Mermin family from Alice to Bob-Charlie in one sided device independent scenario with hidden variable of dimension $2$ having different deterministic distributions at Alice's side} \label{a3}

let us try to generate a LHV-LHS decomposition of the bi-unsteerable noisy Mermin family from Alice to Bob-Charlie in one sided device independent scenario with hidden variables having dimension $2$ having different deterministic distributions at Alice's side. In this case the bi-unsteerable noisy Mermin family can be decomposed in the following way:
\begin{equation}
P_{MF}^{V} (abc|A_x B_ y C_z) = \sum_{\lambda=0}^{1} r_{\lambda} P_{\lambda} (a|A_x) P(b c |B_y, C_z,  \rho^{\lambda}_{BC}).
\label{2d}
\end{equation}
Here, $r_0=u$, $r_1=v$ ($0 <u<1$, $0 <v<1$, $u+v =1$). Since Alice's strategies are deterministic, the two probability distributions $P_{0} (a|A_x)$ and $P_{1} (a|A_x)$ must be equal to any two among $P_D^{00}$, $P_D^{01}$, $P_D^{10}$ and $P_D^{11}$. In order to satisfy the marginal probabilities for Alice, the only two possible choices of $P_{0} (a|A_x)$ and $P_{1} (a|A_x)$ are:\\
1) $P_D^{00}$ and $P_D^{01}$ with $u=v=\frac{1}{2}$\\
2) $P_D^{10}$ and $P_D^{11}$ with $u=v=\frac{1}{2}$.

In case of the first choice, let us assume that 
$P_{0} (a|A_x) = P_D^{00}$, $P_{1}(a|A_x) = P_D^{01}$; $P(b c |B_y, C_z,  \rho^{0}_{BC})$ 
and $P(b c |B_y, C_z,  \rho^{1}_{BC})$ are given by,

\begin{center}
$P(b c |B_y, C_z,  \rho^{0}_{BC})  := \begin{pmatrix}
u_{11} && u_{12} && u_{13} && u_{14}\\
u_{21} && u_{22} && u_{23} && u_{24} \\
u_{31} && u_{32} && u_{33} && u_{34}\\
u_{41} && u_{42} && u_{43} && u_{44}\\
\end{pmatrix} $, 
\end{center}

where $0 \leq u_{ij} \leq 1 \forall i,j$, and $ \sum_{j} u_{ij} =1 \forall i$, and let us assume that $P(b c |B_y, C_z,  \rho^{0}_{BC})$ can be reproduced by performing appropriate quantum measurements on quantum state $\rho_{BC}^{0}$.
and

\begin{center}
$P(b c |B_y, C_z,  \rho^{1}_{BC})  := \begin{pmatrix}
w_{11} && w_{12} && w_{13} && w_{14}\\
w_{21} && w_{22} && w_{23} && w_{24} \\
w_{31} && w_{32} && w_{33} && w_{34}\\
w_{41} && w_{42} && w_{43} && w_{44}\\
\end{pmatrix} $,
\end{center}

where $0 \leq w_{ij} \leq 1 \forall i,j$, and $ \sum_{j} w_{ij} =1 \forall i$, and let us assume that $P(b c |B_y, C_z,  \rho^{1}_{BC})$ can be reproduced by performing appropriate quantum measurements on quantum state $\rho_{BC}^{1}$. \\

Now, with this choice, the box $P_{MF}^{V} (abc|A_x B_ y C_z)$ given by the model   (\ref{2d})
has

\begin{align}
P_{MF}^{V}  = &  \bordermatrix{
\frac{abc}{xyz} & 000 & 001 & 010 & 011 & 100 & 101 & 110 & 111 \cr
000 & \frac{u_{11}}{2} &  \frac{u_{12}}{2} &  \frac{u_{13}}{2} &  \frac{u_{14}}{2} &  \frac{w_{11}}{2} &  \frac{w_{12}}{2} &  \frac{w_{13}}{2} &  \frac{w_{14}}{2}\cr
001 & \frac{u_{21}}{2} & \frac{u_{22}}{2} & \frac{u_{23}}{2} & \frac{u_{24}}{2} & \frac{w_{21}}{2} & \frac{w_{22}}{2} & \frac{w_{23}}{2} & \frac{w_{24}}{2} \cr
010 & \frac{u_{31}}{2} & \frac{u_{32}}{2} & \frac{u_{33}}{2} & \frac{u_{34}}{2} & \frac{w_{31}}{2} & \frac{w_{32}}{2} & \frac{w_{33}}{2} & \frac{w_{34}}{2} \cr
011 & \frac{u_{41}}{2} & \frac{u_{42}}{2} & \frac{u_{43}}{2} & \frac{u_{44}}{2} & \frac{w_{41}}{2} & \frac{w_{42}}{2} & \frac{w_{43}}{2} & \frac{w_{44}}{2} \cr 
100 & \frac{u_{11}}{2} &  \frac{u_{12}}{2} &  \frac{u_{13}}{2} &  \frac{u_{14}}{2} &  \frac{w_{11}}{2} &  \frac{w_{12}}{2} &  \frac{w_{13}}{2} &  \frac{w_{14}}{2}\cr
101 & \frac{u_{21}}{2} & \frac{u_{22}}{2} & \frac{u_{23}}{2} & \frac{u_{24}}{2} & \frac{w_{21}}{2} & \frac{w_{22}}{2} & \frac{w_{23}}{2} & \frac{w_{24}}{2} \cr
110 & \frac{u_{31}}{2} & \frac{u_{32}}{2} & \frac{u_{33}}{2} & \frac{u_{34}}{2} & \frac{w_{31}}{2} & \frac{w_{32}}{2} & \frac{w_{33}}{2} & \frac{w_{34}}{2} \cr
111 & \frac{u_{41}}{2} & \frac{u_{42}}{2} & \frac{u_{43}}{2} & \frac{u_{44}}{2} & \frac{w_{41}}{2} & \frac{w_{42}}{2} & \frac{w_{43}}{2} & \frac{w_{44}}{2} }, 
\label{b2}
\end{align}
where each row and column corresponds to a fixed measurement $(xyz)$ and a fixed outcome $(abc)$ respectively. 

From Eq. (\ref{b2}), it can be seen that
\begin{equation}
P_{MF}^{V} (a b c|A_0 B_ y C_z) = P_{MF}^{V} (a b c|A_1 B_ y C_z), \nonumber
\end{equation}
which is not true for the noisy Mermin family as given in Eq.(\ref{MFO})
with $V>0$. Because
in case of noisy Mermin family given by Eq.(\ref{MFO}), 
\begin{equation}
P_{MF}^{V} (a b c|A_0 B_ y C_z)  = \frac{1 + (-1)^{a \oplus b \oplus c \oplus  yz } \delta_{ y \oplus 1,z} V}{8}, \nonumber
\end{equation}
and 
\begin{equation}
P_{MF}^{V} (a b c|A_1 B_ y C_z) = \frac{1 + (-1)^{a \oplus b \oplus c \oplus  y \oplus yz \oplus z} \delta_{y, z} V}{8}. \nonumber
\end{equation}

Hence, in this case,
though the marginal probabilities for Alice are satisfied, 
all the tripartite joint probability distributions $P_{MF}^{V} (a b c|A_1 B_ y C_z)$ are not satisfied simultaneously for $V>0$.

Similarly, in case of the first choice, if we assume that 
$P_{0} (a|A_x) = P_D^{01}$, $P_{1}(a|A_x) = P_D^{00}$, 
then the marginal probabilities for Alice are satisfied, 
but all the tripartite joint probability distributions $P_{MF}^{V} (a b c|A_1 B_ y C_z)$ are not satisfied simultaneously for $V>0$.

Now, in case of the second choice, let us assume that
$P_{0} (a|A_x) = P_D^{10}$, $P_{1} (a|A_x) = P_D^{11}$; 
$P(b c |B_y, C_z,  \rho^{0}_{BC})$ 
and $P(b c |B_y, C_z,  \rho^{1}_{BC})$ are given by,

\begin{center}
$P(b c |B_y, C_z,  \rho^{0}_{BC})  = \begin{pmatrix}
u^{'}_{11} && u^{'}_{12} && u^{'}_{13} && u^{'}_{14}\\
u^{'}_{21} && u^{'}_{22} && u^{'}_{23} && u^{'}_{24} \\
u^{'}_{31} && u^{'}_{32} && u^{'}_{33} && u^{'}_{34}\\
u^{'}_{41} && u^{'}_{42} && u^{'}_{43} && u^{'}_{44}\\
\end{pmatrix} $, 
\end{center}

where $0 \leq u^{'}_{ij} \leq 1 \forall i,j$, and $ \sum_{j} u^{'}_{ij} =1 \forall i$,  and let us assume that $P(b c |B_y, C_z,  \rho^{0}_{BC})$ can be reproduced by performing appropriate quantum measurements on quantum state $\rho_{BC}^{0}$; and 

\begin{center}
$P(b c |B_y, C_z,  \rho^{1}_{BC})  = \begin{pmatrix}
w^{'}_{11} && w^{'}_{12} && w^{'}_{13} && w^{'}_{14}\\
w^{'}_{21} && w^{'}_{22} && w^{'}_{23} && w^{'}_{24} \\
w^{'}_{31} && w^{'}_{32} && w^{'}_{33} && w^{'}_{34}\\
w^{'}_{41} && w^{'}_{42} && w^{'}_{43} && w^{'}_{44}\\
\end{pmatrix} $, 
\end{center}

where $0 \leq w^{'}_{ij} \leq 1 \forall i,j$, and $ \sum_{j} w^{'}_{ij} =1 \forall i$, and let us assume that $P(b c |B_y, C_z,  \rho^{1}_{BC})$ can be reproduced by performing appropriate quantum measurements on quantum state $\rho_{BC}^{1}$.

Now, with this choice, the box $P_{MF}^{V} (a b c|A_x B_ y C_z)$ given by the model (\ref{2d}) 
has,

\begin{align}
P_{MF}^{V} = \bordermatrix{
\frac{abc}{xyz} & 000 & 001 & 010 & 011 & 100 & 101 & 110 & 111 \cr
000 & \frac{u^{'}_{11}}{2} &  \frac{u^{'}_{12}}{2} &  \frac{u^{'}_{13}}{2} &  \frac{u^{'}_{14}}{2} &  \frac{w^{'}_{11}}{2} &  \frac{w^{'}_{12}}{2} &  \frac{w^{'}_{13}}{2} &  \frac{w^{'}_{14}}{2}\cr
001 & \frac{u^{'}_{21}}{2} & \frac{u^{'}_{22}}{2} & \frac{u^{'}_{23}}{2} & \frac{u^{'}_{24}}{2} & \frac{w^{'}_{21}}{2} & \frac{w^{'}_{22}}{2} & \frac{w^{'}_{23}}{2} & \frac{w^{'}_{24}}{2} \cr
010 & \frac{u^{'}_{31}}{2} & \frac{u^{'}_{32}}{2} & \frac{u^{'}_{33}}{2} & \frac{u^{'}_{34}}{2} & \frac{w^{'}_{31}}{2} & \frac{w^{'}_{32}}{2} & \frac{w^{'}_{33}}{2} & \frac{w^{'}_{34}}{2} \cr
011 & \frac{u^{'}_{41}}{2} & \frac{u^{'}_{42}}{2} & \frac{u^{'}_{43}}{2} & \frac{u^{'}_{44}}{2} & \frac{w^{'}_{41}}{2} & \frac{w^{'}_{42}}{2} & \frac{w^{'}_{43}}{2} & \frac{w^{'}_{44}}{2} \cr 
100 & \frac{w^{'}_{11}}{2} &  \frac{w^{'}_{12}}{2} &  \frac{w^{'}_{13}}{2} &  \frac{w^{'}_{14}}{2} & \frac{u^{'}_{11}}{2} &  \frac{u^{'}_{12}}{2} &  \frac{u^{'}_{13}}{2} &  \frac{u^{'}_{14}}{2} \cr
101 & \frac{w^{'}_{21}}{2} & \frac{w^{'}_{22}}{2} & \frac{w^{'}_{23}}{2} & \frac{w^{'}_{24}}{2} & \frac{u^{'}_{21}}{2} & \frac{u^{'}_{22}}{2} & \frac{u^{'}_{23}}{2} & \frac{u^{'}_{24}}{2} \cr
110 & \frac{w^{'}_{31}}{2} & \frac{w^{'}_{32}}{2} & \frac{w^{'}_{33}}{2} & \frac{w^{'}_{34}}{2} & \frac{u^{'}_{31}}{2} & \frac{u^{'}_{32}}{2} & \frac{u^{'}_{33}}{2} & \frac{u^{'}_{34}}{2}  \cr
111 &  \frac{w^{'}_{41}}{2} & \frac{w^{'}_{42}}{2} & \frac{w^{'}_{43}}{2} & \frac{w^{'}_{44}}{2} & \frac{u^{'}_{41}}{2} & \frac{u^{'}_{42}}{2} & \frac{u^{'}_{43}}{2} & \frac{u^{'}_{44}}{2} }.
\label{b22}
\end{align}

From Eq. (\ref{b22}), it can be seen that
\begin{equation}
P_{MF}^{V} (a b c|A_0 B_ y C_z) = P_{MF}^{V} (\Bar{a} b c|A_1 B_ y C_z), \nonumber
\end{equation}
where $\Bar{a} = a \oplus 1$. The above equation is not true for the noisy Mermin family as given in Eq.(\ref{MFO})
with $V>0$. Because
in case of noisy Mermin family given by Eq.(\ref{MFO}), 
\begin{equation}
P_{MF}^{V} (a b c|A_0 B_ y C_z)  = \frac{1 + (-1)^{a \oplus b \oplus c \oplus  yz } \delta_{ y \oplus 1,z} V}{8}, \nonumber
\end{equation}
and 
\begin{equation}
P_{MF}^{V} (\Bar{a} b c|A_1 B_ y C_z) = \frac{1 + (-1)^{a \oplus 1 \oplus b \oplus c \oplus  y \oplus yz \oplus z} \delta_{y, z} V}{8}. \nonumber
\end{equation}

Hence, in this case,
though the marginal probabilities for Alice are satisfied, 
all the tripartite joint probability distributions $P_{MF}^{V} (a b c|A_1 B_ y C_z)$ are not satisfied simultaneously for $V>0$.

Similarly, in case of the second choice, if we assume that $P_{0} (a|A_x) = P_D^{11}$, $P_{1} (a|A_x) = P_D^{10}$, then the marginal probabilities for Alice are satisfied, 
but all the tripartite joint probability distributions $P_{MF}^{V} (a b c|A_1 B_ y C_z)$ are not satisfied simultaneously for $V>0$.

It is, therefore, impossible to have a LHV-LHS decomposition of the bi-unsteerable noisy Mermin family from Alice to Bob-Charlie in one sided device independent scenario with hidden variable of dimension $2$ having different deterministic distributions at Alice's side.

\section{Reproducing noisy Mermin box using $3 \otimes 2 \otimes 2$ quantum system}	 \label{a4}
Consider, the three spatially separated parties (say, Alice, Bob and Charlie) share the following $3 \otimes 2 \otimes 2$ quantum state:
\begin{equation}
\label{qutrt1}
\rho_2 = V | GHZ \rangle \langle GHZ | + (1-V) |2\rangle \langle 2| \otimes \frac{\mathbb{I}_2}{2} \otimes \frac{\mathbb{I}_2}{2},
\end{equation}
where $| GHZ \rangle = \frac{1}{\sqrt{2}} (|000 \rangle + |111\rangle)$; $0 <V \leq 1$; $|0\rangle$, $|1\rangle$ and $|2\rangle$ form an orthonormal basis in the Hilbert space in $\mathcal{C}^3$;  $|0\rangle$ and $|1\rangle$ form an orthonormal basis in the Hilbert space in $\mathcal{C}^2$ (they are eigenvectors of the operator $\sigma_z$); $\mathbb{I}_2 = |0\rangle \langle 0| + |1\rangle \langle 1|$. Now consider the following two dichotomic POVM $E^1 \equiv \{ E_i^1 (i=0,1) | \sum_i E_i^1 = \mathbb{I}, 0 < E_i^1 \leq \mathbb{I} \}$ and $E^2 \equiv \{ E_j^2 (j=0,1) | \sum_j E_j^2 = \mathbb{I}, 0 < E_j^2 \leq \mathbb{I} \}$, where
\begin{center}
$E_0^1 = \begin{pmatrix}
\frac{1}{2} && \frac{-i}{2} && 0 \\
\frac{i}{2} && \frac{1}{2} && 0 \\
0 && 0 && \frac{1}{2} \\
\end{pmatrix}$, and let us assume that the corresponding outcome is $0$,\\

$E_1^1 = \begin{pmatrix}
\frac{1}{2} && \frac{i}{2} && 0 \\
\frac{-i}{2} && \frac{1}{2} && 0 \\
0 && 0 && \frac{1}{2} \\
\end{pmatrix}$, and let us assume that the corresponding outcome is $1$.
\end{center}

On the other hand,

\begin{center}
$E_0^2 = \begin{pmatrix}
\frac{1}{2} && -\frac{1}{2} && 0 \\
-\frac{1}{2} && \frac{1}{2} && 0 \\
0 && 0 && \frac{1}{2} \\
\end{pmatrix}$, and let us assume that the corresponding outcome is $0$, \\

$E_1^2 = \begin{pmatrix}
\frac{1}{2} && \frac{1}{2} && 0 \\
\frac{1}{2} && \frac{1}{2} && 0 \\
0 && 0 && \frac{1}{2} \\
\end{pmatrix}$, and let us assume that the corresponding outcome is $1$,
\end{center}

Here, matrix form of $E_0^1$, $E_1^1$, $E_0^2$ and $E_1^2$ are written in the basis $\{ |0\rangle, |1\rangle, |2\rangle \}$. Now if  Alice performs the POVMs corresponding to $A_0 = E^1$ and $A_1 = E^2$; Bob performs the projective qubit measurements corresponding to the operators: $B_0 = \sigma_y$ and $B_1 = -\sigma_x$; and Charlie performs the projective qubit measurements corresponding to the operators: $C_0 = \sigma_y$ and $C_1 = -\sigma_x$, then the noisy Mermin family can be reproduced.

\end{document}